\documentclass[sigconf,nonacm]{acmart}

\usepackage{pvldb}
\renewcommand\vldbdoi{XX.XX/XXX.XX}
\renewcommand\vldbpages{XXX-XXX}
\renewcommand\vldbavailabilityurl{}

\usepackage{tikz}
\usetikzlibrary{shapes.geometric}
\usepackage{amsmath}

\usepackage{filecontents}

\usepackage{booktabs}
\usepackage{multirow}
\usepackage{tikz}
\usepackage{amsfonts}
\usepackage{xcolor}
\usepackage{xspace}
\usepackage{tikz}
\usepackage{pifont}
\usepackage[english]{babel}
\usepackage{blindtext}
\usepackage{lipsum}
\usepackage{colortbl} 
\usepackage[ruled,vlined,linesnumbered]{algorithm2e}
\usepackage{booktabs}

\usepackage{filecontents}
\usepackage{xspace}
\usepackage{xcolor}
\usepackage{hhline}
\usepackage{multirow}
\usepackage{soul}

\usepackage{array}
\usepackage{comment}
\usepackage{listings}
\usepackage{dsfont}
\usepackage{algorithmic}
\usepackage{xurl}
\usepackage{setspace}
\usepackage{graphicx}
\usepackage{caption}
\usepackage{ltablex}
\usepackage{subcaption}
\usepackage{amsmath}

\usepackage{booktabs}
\usepackage{makecell}
\usepackage{xcolor}
\usepackage{pifont}
\newcommand{\cmark}{\textcolor{green!60!black}{\ding{51}}}
\newcommand{\xmark}{\textcolor{red!70!black}{\ding{55}}}

\usepackage{ifluatex}

\usepackage{outlines}

\usepackage{xspace}
\usepackage{cleveref}
\usepackage{enumitem}
\newcommand{\name}{Continnum\xspace}
\newcommand{\company}{Company A\xspace}

\definecolor{darkkhaki}{RGB}{189,183,107}
\definecolor{teal}{RGB}{0,128,128}

\newcommand{\tighttablecaption}[1]{\caption{#1}}
\newcommand{\tightcaption}[1]{\caption{#1}}
\newcommand{\tightsection}[1]{\section{#1}}
\newcommand{\tightsubsection}[1]{\subsection{#1}}

\newcommand{\mypara}[1]{\vspace{0.2cm}\noindent{\bf {#1}:}~}
\newenvironment{packeditemize}{\begin{list}{$\bullet$}{\setlength{\itemsep}{0.5pt}\addtolength{\labelwidth}{-4pt}\setlength{\leftmargin}{2ex}\setlength{\listparindent}{\parindent}\setlength{\parsep}{1pt}\setlength{\topsep}{2pt}}}{\end{list}}

\Crefname{section}{Sec.}{Sec.}
\Crefname{algorithm}{Alg.}{Alg.}
\Crefname{figure}{Fig.}{Fig.}

\title{\name: Efficient and Robust Multi-Turn LLM Agent Scheduling with KV Cache Time-to-Live}

\author{Hanchen Li}
\authornote{The first three authors contributed equally to this work.}
\affiliation{%
  \institution{UC Berkeley}
  \city{Berkeley}
  \state{California}
  \country{United States}
}
\email{lihanc@berkeley.edu}

\author{Runyuan He}
\authornotemark[1]
\affiliation{%
  \institution{UC Berkeley}
  \city{Berkeley}
  \state{California}
  \country{United States}
}
\email{runyuan@berkeley.edu}

\author{Qiuyang Mang}
\authornotemark[1]
\affiliation{%
  \institution{UC Berkeley}
  \city{Berkeley}
  \state{California}
  \country{United States}
}
\email{qmang@berkeley.edu}

\author{Qizheng Zhang}
\affiliation{%
  \institution{Stanford University}
  \city{Stanford}
  \state{California}
  \country{United States}
}
\email{qizhengz@stanford.edu}

\author{Huanzhi Mao}
\affiliation{%
  \institution{UC Berkeley}
  \city{Berkeley}
  \state{California}
  \country{United States}
}
\email{huanzhimao@berkeley.edu}

\author{Xiaokun Chen}
\affiliation{%
  \institution{Stanford University}
  \city{Stanford}
  \state{California}
  \country{United States}
}
\email{xiaokunc@stanford.edu}

\author{Hangrui Zhou}
\affiliation{%
  \institution{Tsinghua University}
  \city{Beijing}
  \country{China}
}
\email{zhouhr23@mails.tsinghua.edu.cn}

\author{Huanchen Zhang}
\affiliation{%
  \institution{Tsinghua University}
  \city{Beijing}
  \country{China}
}
\email{huanchen@tsinghua.edu.cn}

\author{Alvin Cheung}
\affiliation{%
  \institution{UC Berkeley}
  \city{Berkeley}
  \state{California}
  \country{United States}
}
\email{akcheung@cs.berkeley.edu}

\author{Joseph Gonzalez}
\affiliation{%
  \institution{UC Berkeley}
  \city{Berkeley}
  \state{California}
  \country{United States}
}
\email{jegonzal@berkeley.edu}

\author{Ion Stoica}
\affiliation{%
  \institution{UC Berkeley}
  \city{Berkeley}
  \state{California}
  \country{United States}
}
\email{istoica@berkeley.edu}

\begin{document}

\begin{abstract}
KV cache management is essential for efficient LLM inference. To maximize utilization, existing inference engines evict finished requests' KV cache if new requests are waiting. 
This policy breaks for agentic workloads, which interleave LLM calls with tools, introducing pauses that prevent effective KV reuse across turns. 
Since many tool calls have much shorter durations than human response multi-turn chatbot, it would be promising to retain the KV cache in during these tools.
However, many challenges remain.
First, we need to consider both the potential cost of recomputation or reloading (if offloading enabled) as well as  the increasing queueing delays after eviction from GPU.
Second, due to the internal variance of tool call durations, the method needs to remain robust under limited predictability of tool call durations.

We present \name, a serving system to optimize job completion time for multi-turn agent workloads 
by introducing time-to-live mechanism for KV cache retention.
For requests that generate tool calls,
\name selectively pins the KV cache in GPU memory with a time-to-live value determined by the reload cost and potential queueing delay induced by eviction. 
When the TTL expires, the KV cache can be automatically evicted to free up GPU memory, providing robust performance under edge cases.
When combined with program-level first-come-first-serve,
\name preserves multi-turn continuity, and reduces delay for agentic workflows.
Evaluations on real-world agents (SWE-Bench, BFCL, OpenHand) with Llama-3.1 8B/70B, Gemma-3 12B, and GLM-4.5 355B shows that \name improves the average job completion times by over 8x while improving throughput. 
\end{abstract}

\maketitle

\vldbtopmatter

\section{Introduction}\label{sec:intro}

KV Cache management is key to large language model inference, impacting both the input processing (prefill) and output generation (decoding) stages~\cite{kwon2023efficient, zheng2024sglang, cheng2025lmcache}.
A critical component of KV cache management is the eviction policy. 
Ideally, the system should avoid evicting tokens that will be referenced in the immediate future. 
Similar to traditional caching systems, 
Existing inference engines assume that KV caches are less important once decoding is finished. To maximize utilization, these engines discard the KV caches of completed requests when new requests are waiting in the queue. We refer to this type of policy \textbf{end-of-turn eviction}.

While end-of-turn eviction works well for multi-turn chat applications, it can significantly degrade the performance of modern agentic workloads, particularly those involving tool calling. 
These agentic applications have become increasingly popular across domains such as software engineering~\cite{yang2024swe}, computer use~\cite{anthropic2024computeruse}, and scientific research~\cite{ren2025towards}.
These workloads characteristically interleave (a) inference steps to derive the next action, and 
(b) execution steps where the agent calls an external tool. 
The output of the tool is subsequently appended to the request context, and a new inference step is initiated in the inference engine. Since the tool call can be much faster (\emph{i.e.,} $\leq 2$s) than human typing speed, this new workload requires changes to end-of-turn eviction.

\begin{figure}[t]
    \centering
    \includegraphics[width=\linewidth]{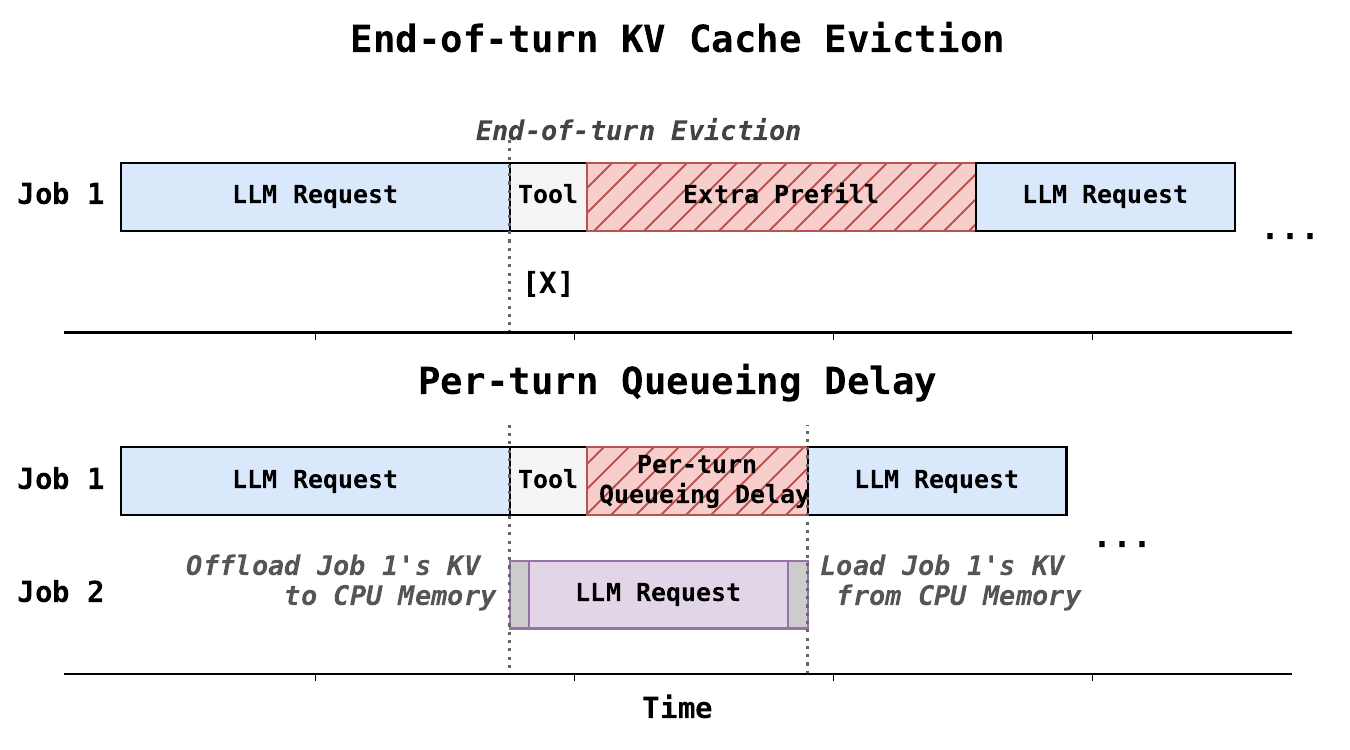}
    \vspace{-6pt}
    \tightcaption{Two main failure modes of prior agent-serving systems. Red blocks represent overhead from suboptimal scheduling and KV-cache management: even with CPU offloading, agents still suffer queueing delay after KV-cache eviction.}%Two main failure modes of previous work on agent serving. Red blocks represent overhead caused by suboptimal scheduling and KV cache management: Even with CPU offloading, agents still suffer from queueing delay after KV cache eviction.} %\lhc{need to specify x and y axis}}
    % \vpsace{4pt}
    \vspace{-4pt}
    \label{fig:intro_motivation}
\end{figure}

The core issue arises after the request's KV cache is evicted 
when the agent transforms from inference step to tool call.
If the KV cache was evicted for this step, 
the engine must recompute the prefix (prefill) or reload from CPU (if CPU offloading is enabled~\cite{cheng2025lmcache}) 
when the tool execution completes and the next inference step begins.
This repetitive prefill introduces substantial delays and reduces overall system throughput. 
More importantly, 
even when CPU offloading is enabled to reuse KV cache, 
eviction causes another problem: \textbf{per-turn queueing delay}. 
When the next inference step has its KV cache evicted from GPU memory, 
even if the KV cache can be reloaded from CPU,
it will also have to wait in the waiting queue for other requests to free up GPU memory before starting inference. 
This per-turn queueing delay can accumulate and result in increasing delay for each agentic program as illustrated in Figure~\ref{fig:intro_motivation}. Since this delay is not measurable by offline profiling, 
we need to design a new model to include its impact.
Moreover, since tool calls can be inherently variable, 
we need to set a maximum KV cache retention time to prevent infinitely long waiting. However, if this time expires just before the tool call, the previous waiting time will be wasted. Thus, we need to carefully set the KV cache retention time to best adapt to the workload.

Previous work fails to address these challenges. 
InferCept~\cite{abhyankar2024infer} makes its KV preserve decision based solely on the reload cost.
But it does not model the per-turn queueing delay that accumulates over turns, nor have a robust mechanism to handle variable tool call durations.
This makes it impractical for real-world deployment.
As we show later in Section~\ref{sec:evaluation},  InferCept accumulates the queueing penalty over turns, resulting in suboptimal performance. 
Autellix~\cite{luo2025autellix} uses end-of-turn eviction and ignores the importance of KV cache retention in multi-turn agent scheduling. 
Pie~\cite{10.1145/3731569.3764814} exposes interfaces but provides no policy for KV cache retention decisions. 
Ayo~\cite{tan2025towards}, Alto~\cite{santhanam2024alto}, and Parrot~\cite{lin2024parrot} assume static workflows and do not apply to dynamic agents.

To provide an efficient and robust solution, we present \name, a serving system that utilizes KV cache time-to-live technique to improve job completion time for multi-turn agent workloads.  
Inspired by previous caching papers, \name introduces a KV cache time-to-live (TTL) mechanism to retain KV cache inside GPU after request finishes to over-ride original end-of-turn evictions.
For each LLM request that generates a tool call during the inference step,
\name models both the prefill/reload cost and the per-turn queueing delay reduction brought retaining KV cache.
After obtaining the benefits of a potential hit based on the above two factors and tool call distributions, \name compares this with the cost of occupying GPU memory space during the TTL time to decide how long the KV cache can stay in GPU memory before being automatically evicted.
This allows the next request to immediately resume if the tool call returns within the TTL window to save prefill and queueing delay.
When the tool call prediction is inaccurate and the tool call takes longer than expected,
\name can correct the mistake robustly by evicting the KV cache after the TTL expires,  preventing severe memory pressure or deadlocks.
Furthermore, \name combines the TTL mechanism with program-level first-come-first-serve scheduling.
This enforces better request ordering and simplifies scheduling for complex agentic workflows.

% \ion{I'd simplify a bit the flow: (1) KVcache management is key to an inference engine prformance both in terms of input and output; (2) Eviction policy is key to KVCahe management. Ideally, we do not want to evict tokens which will be referred immediately in the future. Like with other caching applications, existing inference engines use an LRU policy at the request granularity. (3) While this works well in general, it can hurt the performance of modern agentic workloads, in particular, multi-turn tool calling. These workloads interleave (a) inference for deriving the next action, and (b) using the action to call a tool. The output of the tool is then appended to the request, and a new inference steo is performed. (4) The problem is that if the request was evicted from the KVCache while the tool call was executing, we need to recompute the entire prefix, i.e., we need to prefill the prefix. This adds delays and hurts throughput. (5) This has been observed and previous work, which proposed solution X..... However, [say what are the limitations]. (6) in the paper, we propose Y which addresses this challenge. We do so by [say solution]} 

We implemented \name on top of vLLM with a modular design that can be easily maintained or integrated into other inference engines. \name implemented a tool call handler that is called each time a request enters or leaves the serving engine.
It identifies the tool call, predicts the duration, and decides the timeout of the KV cache pin based on both throughput and request ordering concerns. 
This modular design adds minimal change to the original scheduling logic of the inference engine and allows for future extension to tool-call aware scheduling.
  
To evaluate \name's performance, we conduct extensive experiment on real agentic workloads in function calling~\cite{mao2024bfclv3} and coding agents~\cite{lieret2025miniSWEagent}. 
Across three hardware and model setups, we show that \name reduces delay by 1.12x to 3.66x and improves throughput by 1.10x to 3.22x on multi-turn agentic workloads.
Moreover, we evaluated \name on \company's\footnote{Inference startup anonymized for double-blind} internal testbed and show it can reduce delay for real SWE-agent workloads by up to 8.18x. We will open-source our traces, code, and the agent serving testbed to foster future agent serving research.
% Moreover, we show that \name have robust performance across different DRAM offloading schemes compared with previous methods. \alvin{explain why robustness is important}

In summary, our contributions are the following: 

\begin{packeditemize}
    \item We identify the key cache KV retention problem in agent serving and motivates the need for better solution.
    \item We design \name, a efficient and robust serving system with KV cache time-to-live mechanism to reduce turn-based eviction cost and per-turn  queueing delay.
    \item We demonstrate that \name achieves up to 8.18x improvements in both latency and throughput over previous methods in both emulated and real cases.
    \item  We will open-source our collected agent inference traces, code, and agent serving testbed upon publication.
\end{packeditemize}

\begin{figure*}[t]
    \centering
    \includegraphics[width=\linewidth]{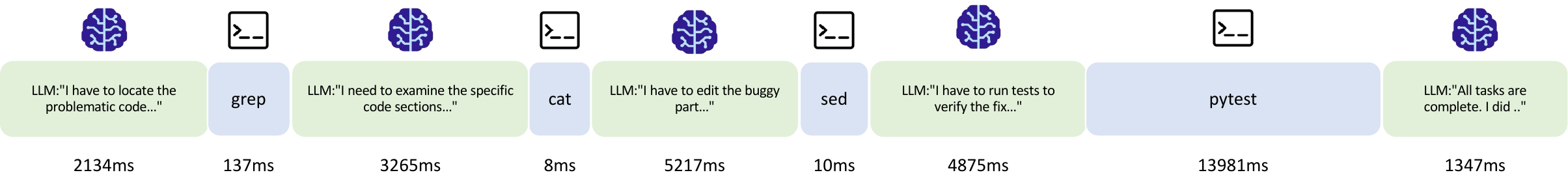}
    \tightcaption{Illustrative example of a SWE-Agent. 
    The agent resolves a software engineering bug step by step with tool calls in the middle. 
    These tool calls have different durations and breaks the continuity of the LLM inference. }
    \label{fig:trace_swe}
\end{figure*}

\section{Background}

\subsection{ReAct Paradigm for Agents}
Most modern agentic workloads follow the \emph{ReAct}-agent loop~\cite{yao2022react}, alternating between a reasoning step where the LLM interprets context and outputs thoughts, and an action step where it invokes external tools.
This paradigm has become the de facto standard: coding agents such as Claude Code~\cite{claude_code_2025} and Cursor~\cite{cursor_agents_2025} adopt it for its clarity and performance, frameworks like LangChain~\cite{langchain_react_agents2025} and LangGraph~\cite{langgraph_state_machines2025} make the pattern broadly accessible, and recent open-weight models including GPT-OSS~\cite{agarwal2025gpt} and Kimi-K2~\cite{kimi-k2-blog} bake tool-call ability directly into the base model.

% Recent work has turned single-turn LLMs into autonomous agents by wrapping them in tool-using workflows. 
% Among these, the ReAct paradigm is especially influential: the agent alternates between internal reasoning (“thought”) and external operations (“action”). 
% Given a query, the model proposes an action plan for this turn; the agent then executes steps by invoking tools (e.g., search engines, calculators), writes the results back into the context, and repeats until it can produce a final answer or take a terminal action.

% Compared with fixed pipelines, this interleaved tool use pattern that gives model the freedom to call tools offers clear benefits. 

% This interleaved reasoning and tool-call paradigm features a few unique advantages in the context of agentic workloads. Thus, the ReAct framework has gradually become the de facto standard of agentic workloads.
% Coding agents including Claude Code~\cite{claude_code_2025} and  Cursor~\cite{cursor_agents_2025} almost all adopt the ReAct workflow for its clarity and performance.
% Frameworks such as LangChain~\cite{langchain_react_agents2025} and LangGraph~\cite{langgraph_state_machines2025} implement ReAct-style agents and graph-based state machines for explicit state to enable more developers build ReAct style agents.

% More importantly, the most recent leading open-source models like GPT-OSS~\cite{agarwal2025gpt} and Kimi-K2~\cite{kimi-k2-blog}  take things further, where it bakes the tool call ability into the base model, and allows for ReAct-style tool calls inside the reasoning chain.
An important trend is that agentic applications increasingly scale this loop into \emph{long-horizon, multi-turn} iterations, repeatedly interleaving thought, tool call, and context update across dozens or even hundreds of turns.
This is reflected in recent benchmarks such as $\tau$-bench for tool-agent-user interaction~\cite{yao2024tau}, MINT for multi-turn tool-augmented interaction~\cite{wang2023mint}, and AgentBench for multi-turn decision-making and tool-use scenarios~\cite{liu2023agentbench}.

\subsection{Limitations of Existing Methods}
% \lhc{talk about infercept}

Previous works fail to handle this emerging complex workload due to three main reasons:

\mypara{Fixed Workflow}
% \qz{This is too much to digest, and our message is not sharp enough. Let's highlight 2-3 limitations of prior work in an organized way, e.g. the first one would be ``reliance on pre-defined, static computation graph". }
One line of work focused on scheduling agentic workflows with \textbf{pre-defined, static} computation graphs.
Teola~\cite{tan2025towards} decomposes applications into primitive-level dataflow graphs and then applies graph-level optimizations. % It applies graph-level optimizations (parallelism, pipelining) to reduce end-to-end latency.
Alto~\cite{santhanam2024alto} focuses on streaming and pipelined execution across distributed components. 
% It tracks partial outputs and routes them hierarchically through compound pipelines. 
Parrot~\cite{lin2024parrot} exposes application-level context to LLM services through Semantic Variables, enabling the engine to infer data dependencies across consecutive LLM requests.
One shared limitation of Teola, Parrot, and Alto is that they all assume static or deterministically defined DAGs and \textbf{could not work with dynamic agent workloads} like ReAct-styled ones whose dependency graphs evolve at runtime. 
This limits these work from optimizing for the wide variety of agents in practice~\cite{anysphere2024cursor, lieret2025miniSWEagent,yan2024bfcl}.
% \alvin{need to justify why dynamic workloads are important. do they frequently happen in practice? do you have stats / refs for that? }

\mypara{No Tool-Call Awareness}
Wadlom et al.~\cite{wadlom2026efficient} study efficient LLM serving for agentic workflows from a data-systems perspective.
Autellix~\cite{luo2025autellix} introduces Program-Level Attained Service (PLAS) scheduling that prioritizes requests with less cumulative service time of the agentic program. 
Tempo~\cite{zhang2025tempo} proposes a scheduler to satisfy the SLOs when facing different types of requests (chat, agent, reasoning), while our focus is particularly on agentic workloads with many-turn and variable tool calls.
Apt-Serve~\cite{gao2025apt} jointly optimizes hybrid-cache management and request scheduling to improve effective throughput, but it targets individual LLM requests rather than dynamic multi-turn agent workflows.
AlignedServe explores prefix-aware batching for LLM serving~\cite{bai2026alignedserve}.
These work fail to consider the unique characteristics of tool calls in agentic workloads, such as their variable durations and the impact on KV cache management.
This oversight can lead to suboptimal scheduling decisions and increased latency, as we demonstrate later in Sec~\ref{sec:motivation-failure}.

\mypara{Insufficient KV Cache Retention Strategies}
Some previous work observed the challenge of KV cache reuse for agent workloads.
Cache-Craft~\cite{agarwal2025cache} manages reusable chunk caches for efficient retrieval-augmented generation, but it focuses on document chunks rather than retaining request-specific agent state across tool calls.
InferCept~\cite{abhyankar2024infer} introduces a ``preserve'' operation that pins the KV cache between tool calls. However, their policy overlooks the multi-turn nature of requests. When KV cache is evicted between turn, this will cause additional queueing time per turn for the program when they come back. In multi-turn scenarios, the queueing time can accumulate for each turn. Ignoring such effects makes them not preserve KV cache in GPU even when there are significant benefits.
Moreover, their preserve operation is fixed and could not adapt to tool use in real time. If the actual tool call time is much longer than predicted, blindly "preserving" the KV cache can cause significant inefficiency.
This makes it impractical for real-world deployment. 
Pie~\cite{10.1145/3731569.3764814} introduces a programmable serving system that decomposes the generation loop into fine-grained handlers. 
It delegates control to user programs, allowing for custom tool call handling. 
However, it requires developers to manually design scheduling for each agent.
and provides no actual method to adapt to dynamic tool-call latencies or multi-turn dependencies.
% However, it does not consider the adverse effect of swapping on the continual execution of the agentic program and will extensively swap KV cache into DRAM. 
% When the next request is sent to the inference engine after tool call, it needs to wait a time period for the GPU memory to be freed by other requests. We refer to these time periods as \textbf{scheduling bubbles}.
% % These scheduling bubbles increases latency for each request, especially when there are many turns in the program. An illustrative example is shown in ~\cref{sec:motivation-failure}.

% However, their KV cache retainment policies have many limitations. First, 

% We demonstrate our improvement over these KV cache preservation techniques in ~\cref{sec:evaluation}.

% While offering great flexibility, its performance relies on how well developers can optimize their inferlets, and our work aims to further improve performance on top of it.

\begin{table}[t]
\tightcaption{Continnum comparison with representative baselines.}
\label{tab:conceptual-comparison}
\centering
\small
\setlength{\tabcolsep}{4pt}
\renewcommand{\arraystretch}{1.0}
\begin{tabular}{lccc}
\toprule
Method & \makecell[c]{Retains\\KV Cache} & \makecell[c]{Includes Per-Turn\\Queueing Delay} & \makecell[c]{Bounds\\Retention Time} \\
\midrule
vLLM               & \xmark & \xmark & \xmark \\
Autellix           & \xmark & \xmark & \xmark \\
Pie                & \cmark & \xmark & \xmark \\
InferCept          & \cmark & \xmark & \xmark \\
\textbf{Continnum}  & \cmark & \cmark & \cmark \\
\bottomrule
\end{tabular}
\end{table}

\section{Motivation}
\begin{figure}[t]

        \includegraphics[width=\linewidth]{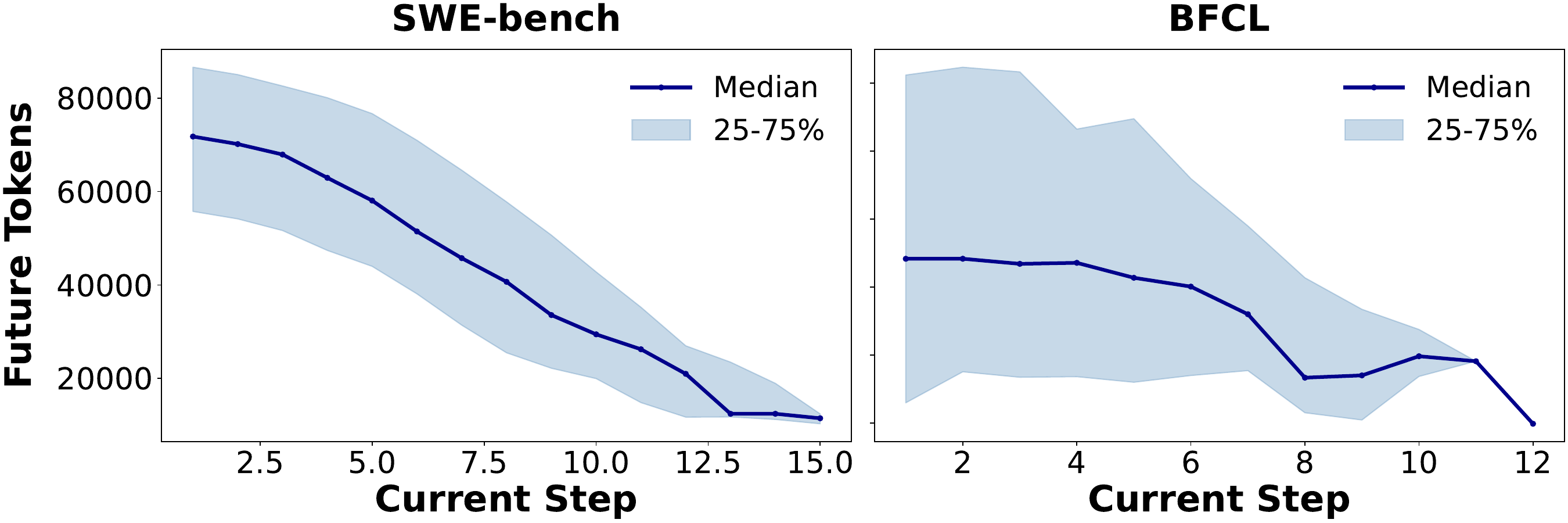}
        
        \label{fig:turn_distribution}
    \tightcaption{Workload characteristics of agentic workloads SWE-Bench and BFCL as used in
   Sec \ref{sec:evaluation}. As the number of steps increase, the requests are closer to finish.}

    \label{fig:workload_char}
\end{figure}

\subsection{Agentic Traces}
\label{sec:motivation-tool}
We begin by analyzing the characteristics of modern agentic workloads.
We collect and analyze 100 mini-swe-agent traces~\cite{lieret2025miniSWEagent} running SWE-Bench~\cite{jimenez2024SWEbench}.
We also collect 100 traces from BFCL V4 Web Search~\cite{mao2025bfclv4web}; both workloads use GPT-5 as the base model.
Figure~\ref{fig:trace_swe} presents an illustrative shortened example trace from SWE-Bench, demonstrating how the agent solves a software engineering task step by step. 

% \lhc{draw a table}
\begin{table}[t]
% \vspace{5pt}
\tighttablecaption{Statistics from two collected datasets. Reported numbers are in the format of (mean, standard deviation).}
\label{tab:dataset}
\centering
\resizebox{\linewidth}{!}{
\begin{tabular}{lccc}
\toprule
\textbf{Dataset}  & \textbf{No. of Turns} & \textbf{Tool Time(ms) } & \textbf{Token Per Program}\\
\midrule
SWE-Bench      & (10.9, 2.1) & (925, 3,550) & (70,126, 19,732) \\
BFCL v4  & (6.3, 2.3) & (1,923, 2,133) & (93,256, 68,687) \\
\bottomrule
\end{tabular}
}
% \vspace{-12pt}
\end{table}

The takeaway is three-fold. First, there are many turns for these novel agentic programs. 
This increase in turn numbers adds additional scheduling difficulty. Second, the tool call times have a varying distribution, but many are short. 
Although the request will be considered finished after these short tool calls are generated,
the next request will arrive soon after the tool call completes, reusing the KV cache.

Last but not least, as shown in Figure~\ref{fig:workload_char}, as the program approaches completion, the expected number of future tokens overall reduces for both workloads. 
This indicates that later turns have shorter expected finish time. 
This suggests that prioritizing requests that came earlier (program-level FCFS) or have executed more turns could be a good approximation 
for the theoretically optimal but clairvoyant shortest remaining time first (SRTF) scheduling policy.
But it is non-trivial to maintain such ordering when tool calls are involved, 
as we will discuss later in \ref{sec:static-kv}.

\begin{figure}[t]
    \centering
    \includegraphics[width=\linewidth]{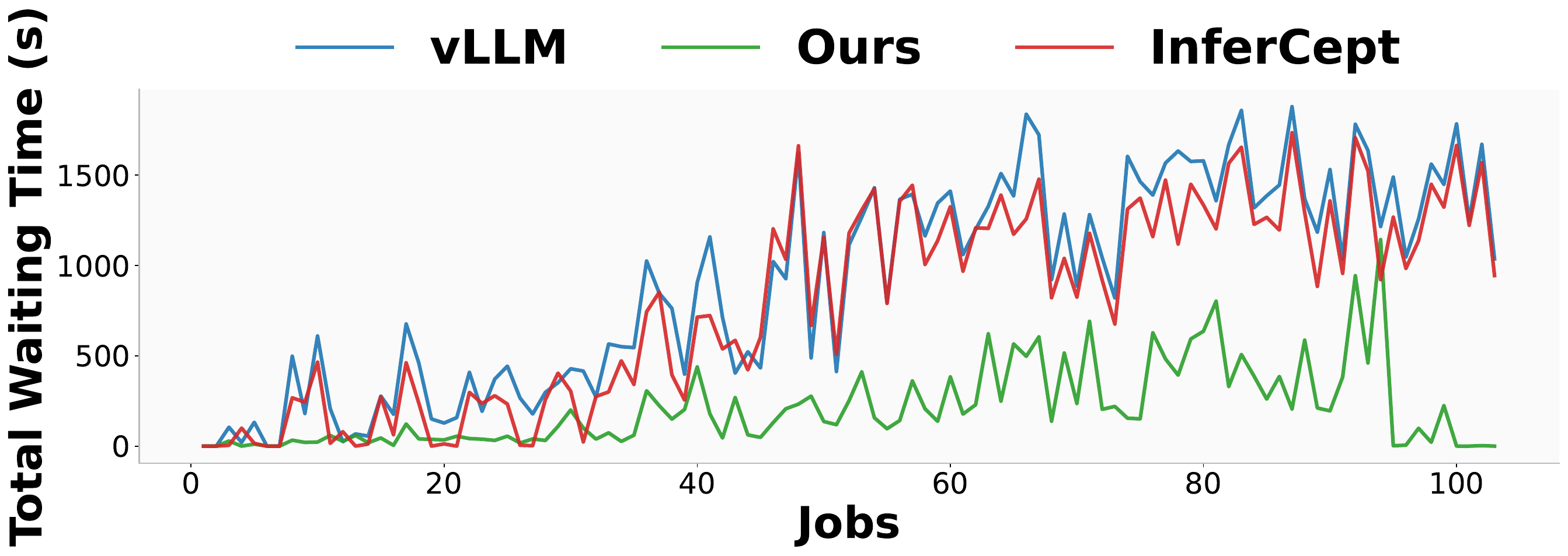}
    \tightcaption{Per-program queueing delay under CPU offloading. InferCept's preserve decision ignores queueing cost, so evicted programs still accumulate substantial waiting time across turns---comparable to vanilla vLLM despite InferCept's reload savings.}
    
    \label{fig:waiting_time_analysis_comparison}
\end{figure}
\begin{figure}[t]

    \centering
    % --- Subfigure 1 ---
    \begin{subfigure}[b]{0.23\textwidth}
        \centering
        \includegraphics[width=\textwidth]{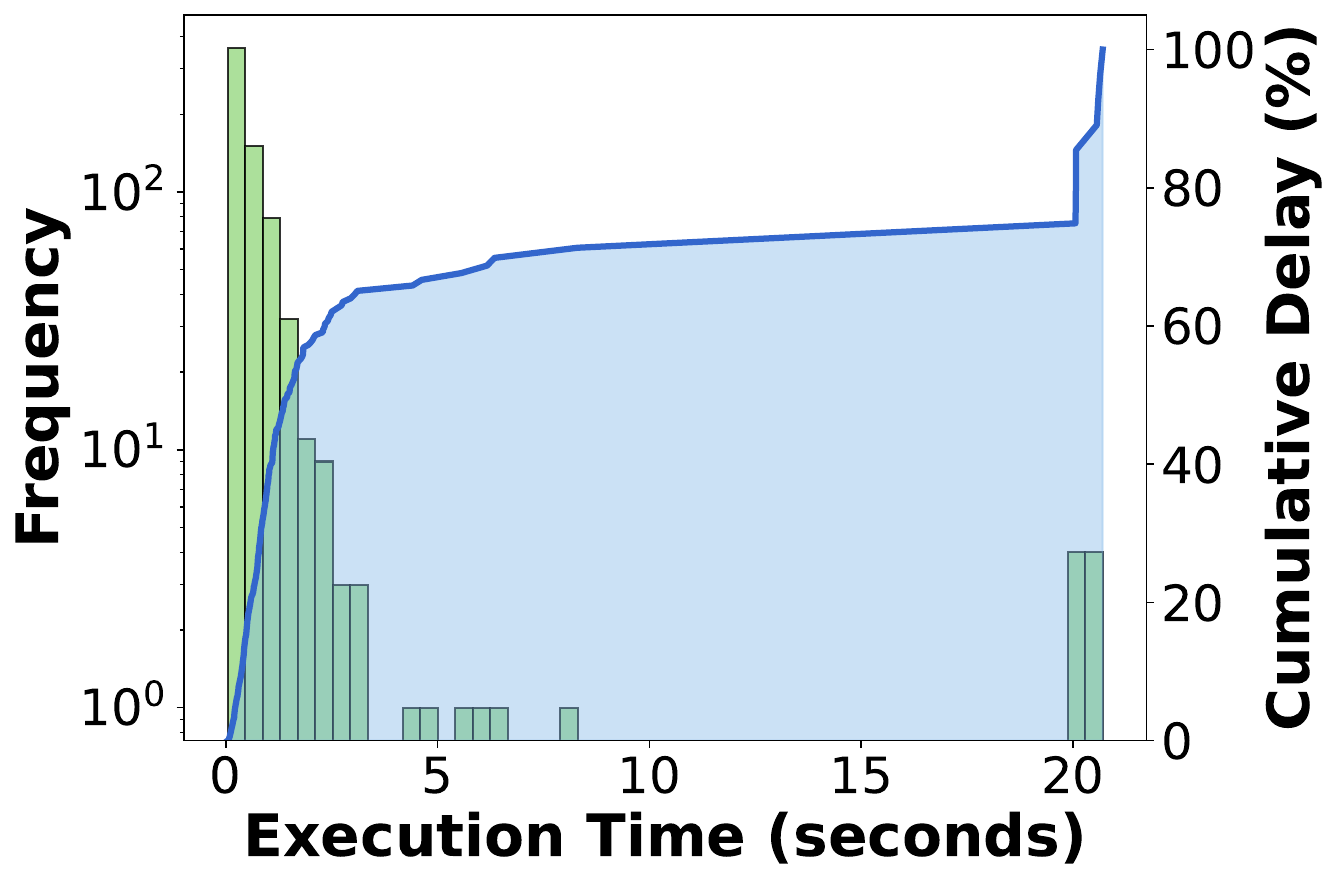}
        \subcaption{BFCL: fetch\_url}  % <-- subtitle
    \end{subfigure}
    \hfill
    % --- Subfigure 2 ---
    \begin{subfigure}[b]{0.23\textwidth}
        \centering
        \includegraphics[width=\textwidth]{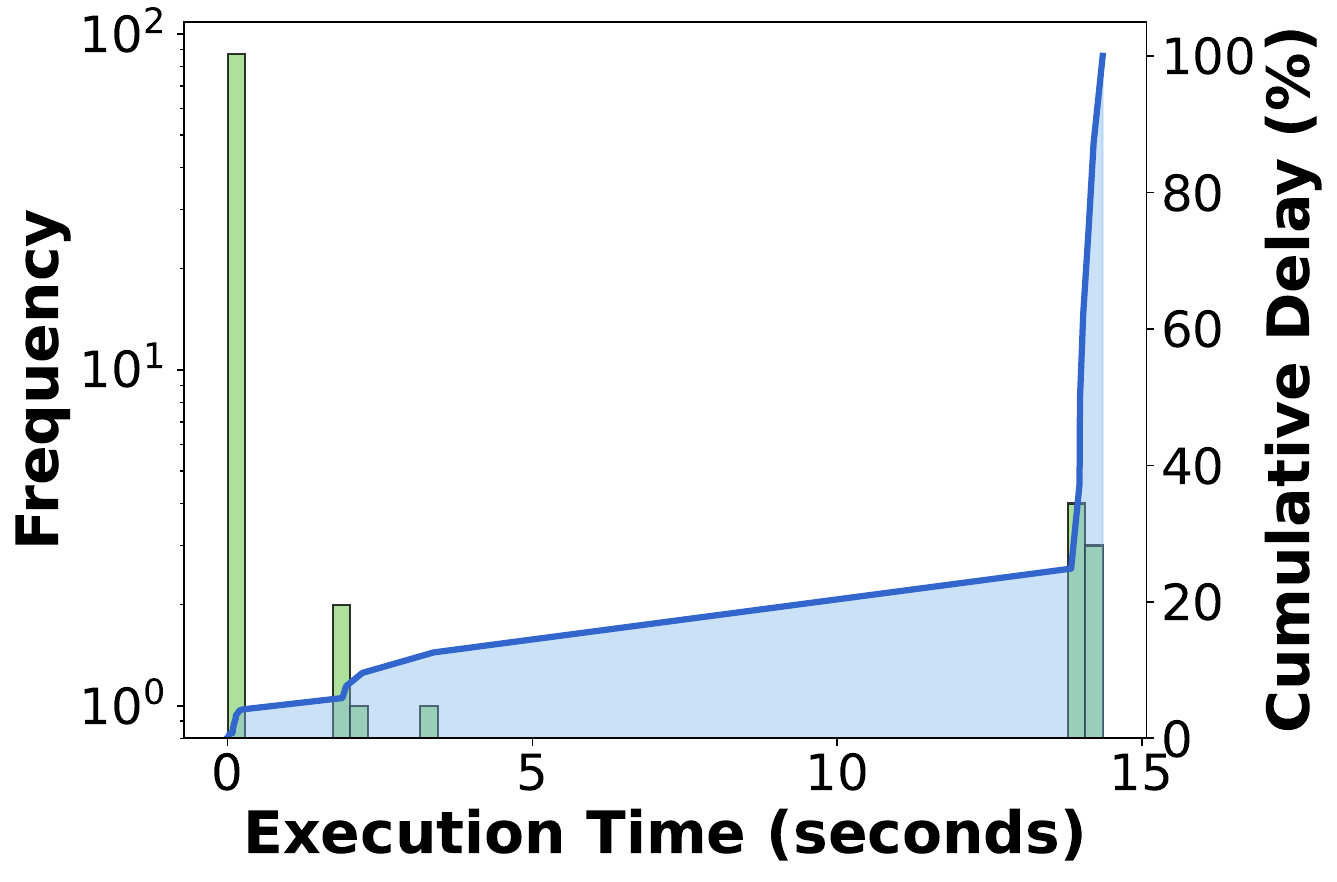}
        \subcaption{SWE-Bench: cd}  % <-- subtitle
    \end{subfigure}

    \tightcaption{Functions' execution time can be extremely long-tailed. Slowest 10\% of fetch\_url account for 52.5\% of the total delay, while slowest 10\% of cd account for 94.1\%. }%
    \label{fig:tool_call_varying}
\end{figure}
\subsection{Challenges for Agentic Workloads}
\label{sec:motivation-failure}
% We will analyze the failure modes of previous works in three directions:
% 1. Using Turn-based Eviction; 2. No consideration Per-Turn Queueing; 3. Static KV retainment that lack robustness under varying tool calls.

\mypara{Turn-based Eviction}
% \lhc{need a prefill figure modified from 3.1a}
% \lhc{Hanchen, give a graph on vllm eviction is a failure for agents, even with previous program level scheduling}
Although these tool calls can be short, inference engines treat them as homogeneous gaps between LLM requests.
vLLM or SGLang will evict a request's KV cache as soon as decoding finishes, implicitly assuming the request is complete. 
However, if the KV cache has been evicted, the engine must either redo the full prefill or reload the KV cache from DRAM when offloading is enabled, incurring additional delay. 
Most systems fall short in handling these scenarios efficiently.

Figure~\ref{fig:intro_motivation} illustrates this effect: 
the tool call creates a pause that triggers KV cache eviction, leading to prefill or KV reload on return.
Thus, it is important to have a KV cache retention policy that considers tool calls to avoid such overheads.

\mypara{Per-Turn Queueing Delay}
\label{sec:queueing}
% \lhc{refine before arxiv}
% \lhc{hanchen, give a graph illustrating the bubbles effect}
The multi-turn nature agent programs also introduces a new challenge for scheduler 
that prior work have critically overlooked.
While the current agent program is waiting on the tool, 
if the scheduler allocates the GPU memory to other requests to maximize throughput, 
the KV cache for the current program will be removed from GPU memory.
When the program's tool call returns and the following LLM request is sent to the scheduler, 
it must wait behind ongoing prefill/decoding of other requests for free GPU space.

This waiting period produces a gap in the execution of the agent program regardless whether the KV cache
 is stored in a CPU DRAM location. 
As shown by Figure \ref{fig:intro_motivation}, this gap also contributes to the delay induced by the tool call besides the previous prefill/loading cost,
 accumulating over turns and causing substantial delays for each program.
Moreover, it also breaks the continuity of the program execution and schedules requests with earlier arrival times after later ones.
Notice that even if we give the highest priority to the new request in the waiting queue, 
it still will be blocked by the ongoing computation of the other requests already in GPU.

Existing works do not consider per-turn queueing delay in their retention policies.
InferCept~\cite{abhyankar2024infer}'s KV ``preserve'' operation is invoked
only when the CPU offloading cost exceeds the estimated GPU occupation cost during the tool call.
Crucially, this decision only accounts for the \emph{reload cost} of the immediate next turn---it entirely ignores the queueing delay that an evicted program will experience when it re-enters the waiting queue behind other active requests.
With fast asynchronous CPU offloading provided by engines like LMCache~\cite{cheng2025lmcache},
the reload cost becomes small, so InferCept's preserve operation is rarely invoked.
Yet the queueing delay persists regardless of offloading speed: even with instant KV reload, the returning request must still wait for GPU memory occupied by other requests to be freed.
Since this queueing cost is incurred at \emph{every} turn, the total accumulated delay grows proportionally with the number of turns per program---precisely the regime where agentic workloads operate.

We demonstrate the performance degradation brought by this lack of consideration for multi-turn scheduling in Figure~\ref{fig:waiting_time_analysis_comparison}.
We profile the total eviction overhead experienced by each request for vanilla vLLM and the InferCept algorithm.
The x-axis represents each agentic program in order of arrival time, while the y-axis denotes the total bubble time for each agentic job
— the total idle period a request experiences in the waiting queue before execution.
Even with InferCept's KV retention, bubbles still persist and causes delay increase despite its throughput improvement over vLLM.

\mypara{Variable Tool Call}
\label{sec:static-kv}
% \lhc{runyuan, give func call variation graphs similar to fetch\_url}
% \lhc{need to draw a cut off point}
Current KV cache retention policy also fail under greatly varying tool calls.
For example, InferCept's approach pins the KV cache in GPU memory until the next request arrives after a tool call.
This methods works fine under stable tool call latencies. 
However, as shown in Figure~\ref{fig:tool_call_varying}, many tool calls exhibit high variability in execution time.
When the tool call takes much longer than expected, the pinned KV cache could occupy GPU memory for a long time.
Similar patterns are observed in database agents,  as external tool calls are more complex.
This leads to inefficient memory usage and even potential deadlocks 
when retained KV cache fully occupies the GPU. 
Thus, a static retention policy lacks robustness in practical scenarios.

% \begin{figure}[t!]
%     \centering
%     \begin{subfigure}[b]{0.95\linewidth}
%         \centering
%         \includegraphics[width=\linewidth]{figures/motivation_ex_short.pdf}
%         \caption{Pin time too short: KV cache gets evicted before tool call completes, requiring expensive recomputation or still causing bubbles.}
%         \label{fig:motivation_ex_bad_pin_a}
%     \end{subfigure}
%     \hfill
%     \begin{subfigure}[b]{0.95\linewidth}
%         \centering
%         \includegraphics[width=\linewidth]{figures/motivation_ex_bad_pin.pdf}
%         \caption{Pin time too long: GPU memory becomes occupied, blocking other requests and reducing overall throughput.}
%         \label{fig:motivation_ex_bad_pin_b}
%     \end{subfigure}
    
%     \tightcaption{Improper pin time can cause performance degradation. We need to set TTL smartly.}
%     \label{fig:motivation_ex_bad_pin}
% \end{figure}

\section{\name Scheduling Algorithm}
% \lhc{edit again tuesday night}
% \paragraph{Opportunity with KV Cache Time to Live}
\label{sec:motivation-pin}
% \lhc{add figure showing good and bad TTL time}
Given the failure of previous work, we identify the key question in serving agentic workloads: How to efficiently and robustly retain KV cache in multi-turn scenarios?

An optimal KV cache retention policy should include the following features:
\begin{packeditemize}
    \item It should retain KV cache for requests that will reuse them soon after tool calls, minimizing prefill/loading overheads.
    \item It should consider the multi-turn continuity of agent programs, reducing waiting and preserving program order.
    \item It should be robust to varying tool call latencies.   
\end{packeditemize}
\begin{figure}[t]
    \centering
    \includegraphics[width=\linewidth]{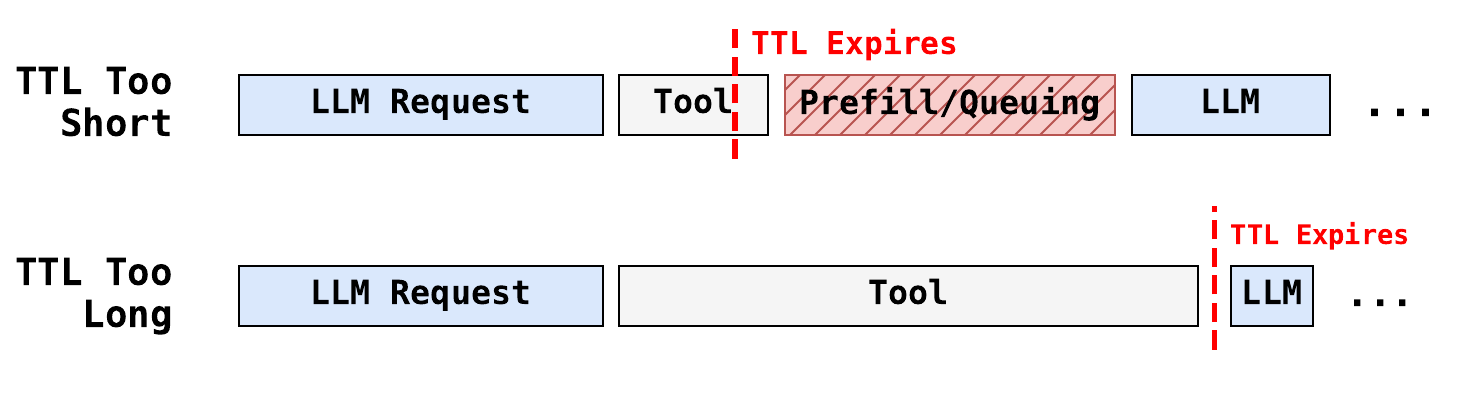}
    \tightcaption{Time-to-live needs to be well set to balance between memory usage and the prefill plus per-turn queueing delay.} %\lhc{need to specify x and y axis}}
    \label{fig:ttl_challenge}
\end{figure}
In order to achieve the robustness guarantee, we propose to borrow the idea of Time-to-live (TTL) from traditional systems:
for each request's KV cache, we give a TTL value to define the maximum duration for it to remain in GPU memory. 
This prevents long-running or failed tool calls from blocking GPU resources indefinitely while retaining KV cache.

However, setting appropriate TTL values for each KV cache entry is challenging compared with static preserve operations.
First, the TTL value should not be too large. If the timeout duration is too long as shown in Figure \ref{fig:ttl_challenge}, 
the pinned KV cache occupies GPU memory unnecessarily, blocking other requests and reducing overall system throughput. 
On the other hand, if the pin time for the specific KV cache is too short,  
the KV cache is evicted before the tool call completes, still causing expensive recomputation or scheduling bubble 
despite wasted GPU occupation time.

Given these tradeoffs, the TTL value should be set carefully. Only if we can set appropriate TTL values based on based on tool call durations, 
prefill/loading costs, and the measurement to program continuity, 
we can balance the benefit of cache reuse against the need to maintain system throughput for other requests to achieve good performance.

\begin{algorithm}[t!]
    \caption{\name's Scheduling Algorithm}
    \label{alg:main-algo}
    
    \SetKwInOut{GlobalState}{Global state}
    \GlobalState{waiting queue $Q$; TTL map $P$ (records pinned programs and their TTLs); historical tool-call records $S$, where $S[f]$ denotes the recorded tool-call information for tool $f$}
    \BlankLine
    
    \SetKwProg{Fn}{Function}{:}{}
    \SetKwIF{If}{ElseIf}{Else}{If}{then}{else if}{else}{end if}
    \SetKwFor{While}{While}{do}{end while}
    \SetKwFor{ForEach}{For each}{do}{end for}
    
    \Fn{\upshape $\mathsf{OnRequestArrive}\textup{(}\text{request } r\textup{)}$}{
        $Q \leftarrow Q \cup \{r\}$, $id \leftarrow$ Program ID of $r$\;
        \If{$id$ is a seen program}{
            $(f, t) \leftarrow$ Tool-call information from $r$\;
            $S[f] \leftarrow S[f] \cup \{t\}$\; \label{line:update-stats}
        }
    }
    \Fn{\upshape $\mathsf{OnRequestFinish}\textup{(}\text{request } r\textup{)}$}{
        \eIf{$r$ is the last request of its program}{
            Free KV cache used by $r$\;
        }{
            $f \leftarrow$ Next tool to be called after finishing $r$\;
            $id \leftarrow$ Program ID of $r$\;
            $P[id] \leftarrow \mathsf{CalcTTL}(r, S[f])$\; \label{line:calc-timeout}
        }
    }
    \Fn{\upshape $\mathsf{Schedule}\textup{()}$}{ \label{line:schedule-start}
        \While{$Q$ is not empty}{
            \ForEach{$id$ in $P.\mathsf{keys}$}{ \label{line:release-start}
                \If{current time $>$ $P[id]$ and $id \notin Q.\mathsf{programs}$}{
                    Free KV cache used by $id$'s last request\;
                    $P \leftarrow P \setminus (id, P[id])$\;
                }
            } \label{line:release-end}
            
            $r \leftarrow \text{argmax}_{r' \in Q} \ \mathsf{CalcPriority}(r', P)$\; \label{line:priority}
            \eIf{$r$ cannot fit into memory}{ \label{line:break}
                \textbf{break}\;
            }
            {
                $Q \leftarrow Q \setminus \{r\}$\; \label{line:issue-start}
                Issue $r$ to running\;
                $id \leftarrow$ Program ID of $r$\;
                \lIf{$id \in P.\mathsf{keys}$}{
                    $P \leftarrow P \setminus (id, P[id])$
                } \label{line:issue-end}
            }
        }
    }
\end{algorithm}

% \qm{1. pesudocode; 2. line-by-line explanation}
\subsection{Utility Model}
\label{sec:cost-model}
% \qm{i think we need to use fine-grained statistics here like SkyStore.}

\begin{table}[t]
\caption{Key notations in \name's cost model for a request $r$ and its associated tool-call $f$.}
\label{tab:cost-model-notation}
\small
\centering
\resizebox{1\linewidth}{!}{
\begin{tabular}{@{}ll@{}}
\toprule
\textbf{Notation} & \textbf{Description} \\
\midrule
$\tau$ & TTL \\
$\mathsf{MemUsage}(r)$ & GPU memory occupied by $r$ \\
$\mathcal{M}$ & Average memory occupied by the seen requests \\
% $\mathcal{V}$ & GPU-CPU practical bandwidth \\
$\mathsf{CacheMissCost}(r)$ & Cost of reloading $r$ \\
$\mathsf{Prefill\text{-}Reload}(r)$ &  Time for reconstructing KV cache in GPU\\
$\mathsf{OutofOrderCost}(r)$ & Cost of out-of-order for request $r$ \\
$\eta$ & Memoryfulness factor of the workload \\
$\mathcal{T}$ & Average waiting time \\
$\mathcal{P}(\tau, f)$ & Estimated finish-within-TTL probability for $f$\\
\bottomrule
\end{tabular}
}
\end{table}

To set an effective TTL value (in seconds) for pinning a request’s KV cache, \name must choose the value that best balances the benefit of potential reuse against its cost. Both the benefit and the cost are measured in units of time, since they ultimately translate into changes in the total job completion latency across all programs. Mathematically, given a request \(r\) and a TTL value \(\tau\), \name estimates \(\mathsf{Cost}(\tau, r)\) and \(\mathsf{Benefit}(r)\) for pinning the KV cache of request \(r\) for \(\tau\).
For simplicity, \(\mathsf{Benefit}(r)\) assumes that the next request arrives within the TTL window. The case where TTL expires before the tool call returns is addressed in Sec. \ref{sec:pin_logic}.

\vspace{-1ex}
\paragraph{Cost Estimation.}
The cost of pinning a request’s KV cache comes from the opportunity cost of occupying GPU memory that could otherwise be used to serve other requests:
\[
\mathsf{Cost}(\tau, r) = \frac{\mathsf{MemUsage}(r)}{\mathcal{M}} \times \tau,
\]
where \(\mathsf{MemUsage}(r)\) is the amount of GPU memory used by the KV cache of request \(r\), \(\mathcal{M}\) is the average GPU memory footprint of active requests, and \(\tau\) is the TTL value.

The ratio \(\frac{\mathsf{MemUsage}(r)}{\mathcal{M}}\) represents how many average requests are blocked when \(r\) is pinned. In other words, if pinning \(r\) occupies the same memory as \(k\) requests, then pinning \(r\) adds \(\tau\) latency to approximately \(k\) other requests. We assume that the waiting queue always contains enough requests for this blocking effect to occur when KV retention is necessary.

\vspace{-1ex}

\paragraph{Benefit Estimation.}
The benefit of pinning a request's KV cache is realized when the request is re-issued within the TTL period, 
allowing it to avoid the overhead of reloading or prefilling the KV cache from $r$'s program while saving the per-turn queueing delay:
\begin{align*}
    \mathsf{Benefit}(r) &= \mathsf{CacheMissCost}(r) + \mathsf{OutofOrderCost}(r)
\end{align*}

Here, $\mathsf{CacheMissCost}(r)$ measures the cost of reloading or prefilling the KV cache for request $r$ 
and $\mathsf{OutofOrderCost}(r)$ measures the expected queueing delay for the request due to waiting for other requests to free GPU memory. We use the sum of cost prevented as the benefit.

Similar to $\mathsf{Cost}(\tau, r)$, we can measure $\mathsf{CacheMissCost}(r)$ by (1) the context reconstruct overhead $\mathsf{Prefill\text{-}Reload}(r)$; and (2) the approximate number of requests will experience the additional latency overhead $\frac{\mathsf{MemUsage}(r)}{\mathcal{M}}$.
The cost is formally defined as follows:
\[
\mathsf{CacheMissCost}(r) = \frac{\mathsf{MemUsage}(r)\times\mathsf{Prefill\text{-}Reload}(r)}{\mathcal{M}}
\]
$\mathsf{Prefill\text{-}Reload}(r)$ is the time cost for prefill or reloading depending on whether CPU offloading is turned on. This is based on a quick offline profiling described in Sec~\ref{sec:profile}.

\mypara{Measuring the expected queuing delay}
As discussed in Sec.~\ref{sec:queueing}, retaining KV cache also eliminates the queueing delay that a returning program would experience if evicted---even when CPU offloading makes reload itself fast.
This $\mathsf{OutofOrderCost}$ component is the key term absent from prior retention policies such as InferCept~\cite{abhyankar2024infer}, which only considers the reload cost.
By modeling this term, \name can justify retaining KV cache even when reload is cheap, as long as the queueing delay savings outweigh the GPU memory occupation cost.
Note that the queueing delay benefit is closely tied to the memoryfulness of the workload, \emph{i.e.,} whether the number of remaining steps reduces predictably as the program progresses.

For example, if the number of requests issued by each program follows a geometric distribution, then the expected number of remaining requests is constant regardless of how many have already been served; in this case, pinning provides no benefit for the queueing delay since keeping the order does not accelerate finishing short jobs first. In contrast, if each program issues a fixed number of requests, then the TTL can eliminate the queueing cost by approximating Shortest Job First. 

Let $N$ be the total number of requests in a program and $k$ the number of requests that have already been served.
We define the following \emph{memoryfulness factor} 
\[
\eta = -\mathrm{Corr}(k, N - k)\big.
\]
% where $R(k)$ denotes the expected number of remaining requests conditioned on having served $k$ requests.
% \begin{align*}
%     R(k) = \mathbb{E}[ N - k \mid k].
% \end{align*}
We can see this factor models the degree of memoryfulness in the workload well: when the workload is fully memoryless, we have that $k$ is independent to $N - k$, leading to $\eta = 0$. Conversely, when the workload is fully memoryful, \emph{i.e.,} all programs have the same fixed number of requests, we have $\mathrm{Corr}(k, N - k) = \mathrm{Corr}(k, -k) = -1$, resulting in $\eta = 1$.
Note that, in some cases $\eta$ may be less than zero (extremely long-tail turn distribution), indicating an \emph{anti-memoryful} pattern in which making progress on a program appears to reveal even more remaining work. We did not observe such patterns but \name is designed with such extreme workloads in mind: it would be preferable to serve each program only briefly and switch frequently to adapt to the long-tail turn distribution.
% Our cost model naturally accounts for this behavior.

Now, we are ready to define the $\mathsf{OutofOrderCost}(r)$ based on the $\eta$ above.
When $\eta = 1$, the delay is exactly the waiting time when the program of $r$ returns back to the waiting queue.
To match this, we record the average waiting time per unit context size for the historical requests in this workload as $\frac{\mathcal{T}}{\mathcal{M}}$, where T is the average queueing delay for previous requests.
In this case, the delay can be well measured by $\frac{\mathcal{T}}{\mathcal{M}}\times \mathsf{MemUsage}(r)$.
Here, we consider $\mathsf{MemUsage}(r)$ since large-context requests are harder to schedule (they must wait for enough contiguous memory to be freed).
For the general cases, we define the out-of-order cost as follows:
\[
    \mathsf{OutofOrderCost}(r) =  \frac{\mathcal{T}}{\mathcal{M}}\times \mathsf{MemUsage}(r) \times \eta.
\]

% \paragraph{Normalization.}

% where $\mathsf{ServeCost}(r)$ denotes the cost of serving request $r$, and can also be determined based on hardware bandwidth and the task size of $r$.
% We use this term to approximate the reduction in switching idle: when 
% $r$ is evicted, the resources it occupied are likely to be immediately used by requests from other programs with comparable serving costs, so the idle bubble eliminated by pinning $r$ can be reasonably estimated by its serving time.

% \vspace{-1ex}
% \paragraph{Implementation.}
% To make the above cost-benefit model practical, we implement the following components involved:
% \lhc{this part should go to sec 5. Let me rewrite sec 5}

% \begin{itemize}[leftmargin=*, itemsep=0pt]
%     \item $\mathsf{MemUsage}(r)$: \qm{todo@hanchen}
%     \item $\otimes$: \qm{todo@hanchen}
%     \item $\mathsf{ReloadCost}(r)$: \qm{todo@hanchen}
%     \item $\mathsf{ServeCost}(r)$: \qm{todo@hanchen}
% \end{itemize}

\subsection{Setting the TTL Value}
\label{sec:pin_logic}
% \lhc{need to add simplified version}
In this part, we describe how \name sets the TTL value for KV cache based on the cost-benefit model above and historical tool-call information.
As in Algorithm~\ref{alg:main-algo} (line~\ref{line:calc-timeout}), \name determines the optimal TTL value $\tau^{*}$ to maximize the expected net benefit of retaining the KV cache:
\begin{align}
    \label{formula:argmin}
    \tau^{*} = \text{argmax}_{\tau}\ \mathcal{P}(\tau, f) \times \mathsf{Benefit}(r) - \mathsf{Cost}(\tau, r),
\end{align} where $\mathcal{P}(\tau, f)$ estimates the probability that the tool call $f$ completes within time $\tau$.
This formula captures the expected net benefit, in terms of total job latency, of retaining the KV cache of $r$ for a duration of $\tau$
By eliminating the shared $\frac{\mathsf{MemUsage(r)}}{\mathcal{M}}$, the formula above can be transformed to
\begin{align}
    \label{formula:argmin2}
    \text{argmax}_{\tau}\ \mathcal{P}(\tau, f) \times \big(\mathcal{T}\cdot\eta +\mathsf{Prefill\text{-}Reload(r)}\big) - \tau,
\end{align}
indicating that we only need to additionally compute $\mathcal{T}$ and $\mathcal{P}(\tau, f)$ in our implementation. $\mathcal{T}$ can be estimated as the sliding window average for queueing delay experienced by requests who was evicted.
Since we cannot fully predict the duration of the next tool call, we estimate $\mathcal{P}(\tau, f)$ using the empirical CDF derived from historical tool-call records $S[f]$.
Specifically, we calculate it as the following:
\begin{align*}
\mathcal{P}(\tau, f) = \frac{1}{{|S[f]|}} \cdot {\sum_{t \in S[f]} \mathbb{I}[t \leq \tau]} 
\end{align*},
where $\mathbb{I}[\cdot]$ is the indicator function.
Finally, we solve Equation~\eqref{formula:argmin2} by enumerating all unique tool-call durations 
recorded in $S[f]$ as candidates (including $\tau = 0$) and selecting the one with the highest expected reward.

\vspace{-1ex}
\paragraph{Cold-start Handling.}
When the number of historical records in $S[f]$ is small, the empirical CDF estimation may be unreliable.
In this case, we first try to use the global tool-call information to estimate $\mathcal{P}(\tau, f_{\text{any}})$, 
which can be computed as $\sum_{t \in S} \mathbb{I}[t \leq \tau] / |S|$.

Moreover, at the very beginning of engine serving, even the global records might not be reliable.
To address this, we design a minimal version of \name that uses a fixed TTL threshold $T_{\text{default}}$, derived from the same cost model by assuming that the tool-call duration follows an exponential distribution with unit mean, \emph{i.e.,} $\mathsf{ToolCallDuration} \sim \text{Exp}(1)$; and the workload is fully memoryful, \emph{i.e., } $\eta = 1$.
$T_{\text{default}}$ is then set to the optimal $\tau^{*}$ under this scenario. 

% We leave more discussion for the minimal version in \Cref{sec:simplified-version}.
% \lhc{qiuyang, can you add the below commented part to appendix}

In practice, we set a threshold $M$ to decide whether to use fixed TTL, global records, or the fine-grained estimation above based on $S[f]$.
That is, we use $T_{\text{default}}$ when $|S| \leq K$; otherwise, we use the global records when $|S[f]| \leq K$, and use the fine-grained TTL setting for the remaining cases.
In our implementation, we set $K = 100$ and initialize $\mathcal{T}$ as zero.

Moreover, since agents are usually post-trained with the tools before production~\cite{cao2025skyrlagentefficientrltraining, cheng2025agentr1trainingpowerfulllm, deepswe2025}, users can also obtain these cost-model statistics during training .

% \lhc{agent always have training}

% We also demonstrate the effectiveness of only using this version in our experiments (\Cref{sec:ablation}), showing that even with a simplified pinning strategy, \name can still achieve significant performance improvements over baseline schedulers.

\vspace{-1ex}
% \paragraph{Unpinning.}
% After computing the TTL, we record it in the map $P[id]$ associated with request $r$'s program, where $id$ denotes the program's identifier. We unpin the KV cache from GPU when the TTL has expired.
% This value will serve as a deadline for releasing the pin later.
% When the scheduler runs (line~\ref{line:schedule-start}), it first checks all pinned programs in the pin map.
% If the current time exceeds a program’s pin timeout and the program is not currently in the waiting queue (lines~\ref{line:release-start}–\ref{line:release-end}), the scheduler releases the pin by freeing the occupied KV cache and removing the program from the pin map.

\subsection{Scheduling Priority}
\label{sec:scheduling_priority}
% \qm{1. preempt first; 2. pin second; 3. FCFS}
In order to keep the scheduling compatible with the TTL algorithm, we need to re-define the request priority in inference engines.
\name introduces a TTL-aware priority that elevates pinned requests within TTL to preserve continuity 
while still preserving program-level FCFS ordering.
Specifically, the scheduler assigns each request $r$ in the waiting queue $Q$ a multi-key priority tuple and ranks requests according to the following criteria (in order):
\begin{packeditemize}
% \item \textbf{Able to fit into memory:} Requests that can fit into memory are prioritized over those that cannot, ensuring that the scheduler always issues feasible requests first.
\item \textbf{Preempted status:} Same as the original engine, preempted requests (due to running queue contention) are prioritized over non-preempted ones.
\item \textbf{TTL status:} In other requests, requests retained within the TTL window are prioritized 
over unpinned ones. 
\item \textbf{Program-level arrival order:} Finally, within each category, requests are ordered by their program-level arrival time to maintain FCFS fairness.
\end{packeditemize}

\begin{figure}[t]
\centering
\includegraphics[width=\linewidth]{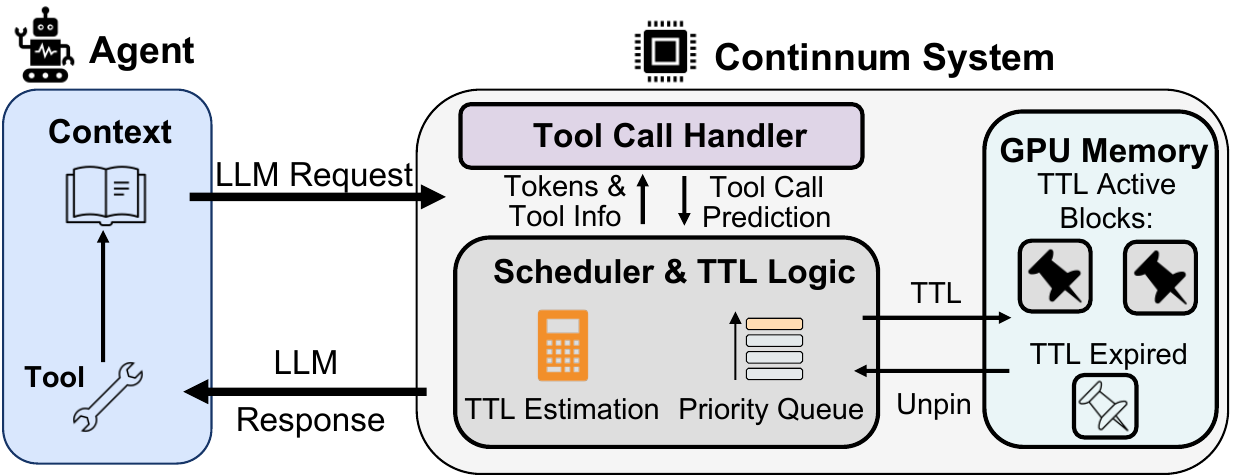}
\tightcaption{System Overview of \name }
\label{fig:system-overview}
\vspace{8pt}
\end{figure}

\section{\name System Design}
% \lhc{add ttl}
% \subsection{System Overview}

In \name, our design goal is a modular architecture that requires minimal changes to the core
inference-engine scheduler loop. 
% Figure \ref{fig:system-overview} illustrates the end-to-end
% workflow. 
On the client side, we attach a program identifier (\texttt{program\_id}) to every inference request so the system can recognize multi-turn agent programs and reason about tool calls across steps.

Upon arrival at the serving engine, requests enter the existing scheduler loop. \name adds a thin Tool-Call Handler that is invoked on request arrival and
completion. The handler parses tool calls from LLM outputs, tracks per-tool latency using observed
inter-request intervals within the same \texttt{program\_id}, and returns TTL to the
scheduler. The scheduler uses this hint to pin the request's KV cache for potential reuse by the
next step, and later unpins it either when the TTL value expires or when the program
terminates. 

\subsection{Tool Call Handler}

The tool call handler is a separate class invoked by the main scheduler after the arrival or at the finish of a request. This decoupled structure ensures that tool handling logic remains isolated from the core scheduling loop, ensuring extensibility for future parsers or tool-aware policies.

% \myparaq{Actions When Request Arrives}
% When a request arrives, the Tool Call Handler will check if the previous request of the program contains a tool call. 
% If there is a tool call in the previous request, the Tool Call Handler will calculate the function call duration based on the difference between the current timestamp and the timestamp that the response was sent back to the client.

% When a request finishes, the Tool Call Handler will identify the tool call, provides a timeout value to the main scheduler loop, and record the request finish time for future tool call time estimation.

\mypara{Identifying the Tool Call}
% \lhc{change to simpler}
When the scheduler completes request, it forwards the response to the tool-call handler, which determines whether the response includes a tool invocation. The handler parses the message according to the function call schema, as the LLM outputs frequently adopt a standardized tool call structure such as the OpenAI schema:
% \vspace{-2ex}
% \begin{verbatim}
% % [{ "id": "fc_12345xyz",
% %    "call_id": "call_12345xyz",
% %    "type": "function_call",
% %    "name": "get_weather",
% %    "arguments": {"location":"Paris"} }]

% \end{verbatim}
% \begin{minted}[fontsize=\small, bgcolor=gray!5, frame=single]{json}
% {
%   "id": "fc_0",
%   "call_id": "call_0",
%   "type": "function_call",
%   "name": "get_weather",
%   "arguments": {"location": "Paris"}
% }
% \end{minted}
\lstset{
  basicstyle=\ttfamily,
  stringstyle=\color{blue!70!black},
  keywordstyle=\color{red!70!black}, % For true/false/null
  showstringspaces=false,
  breaklines=true,
  morekeywords={type, name, id, call\_id, arguments} % Tell it "type" and "name" are special
}
\begin{lstlisting}
{
  "id": "fc_0",
  "call_id": "call_0",
  "type": "function_call",
  "name": "get_weather",
  "arguments": {"location": "Paris"}
}
\end{lstlisting}

For this example schema, the handler checks each returned message block’s \texttt{type}; if it indicates a function/tool call, the handler extracts the call’s \texttt{name} and uses this as the tool call type. 
In SWE-Bench, it is guaranteed that each LLM’s response containing a function call will include exactly one \texttt{bash} function call. We extract the string within the \texttt{bash} block and use the first word afterwards as the tool call name.

More function call format examples for different LLMs~\cite{lin2025overview, qwen_function_calling2025} can be handled by extending this parser with the same schema-specific extraction logic.

\mypara{Recording the tool finish time}
For each LLM request \(i\) in a program identified by a program ID \(p\), the handler records a server-side completion timestamp \(t_{\mathrm{finish}}^{p,i}\) along with tool call name when scheduler records a finished request with tool call output. 
When the next request \(i{+}1\) with the same \(p\) arrives, we observe its server-side arrival timestamp \(t_{\mathrm{arrive}}^{p,i{+}1}\) and compute the inter-request interval \(t_{\mathrm{arrive}}^{p,i{+}1} - t_{\mathrm{finish}}^{p,i}\). 
We record this interval as the execution time of the tool call this time to store for TTL computation in the future.
% The handler persists \(\Delta_{\mathrm{obs}}\) together with the parsed tool name stored . 
%  The estimator provides a robust prediction that downstream components use to set pin timeouts and resume priorities. If no predecessor exists for \(p\), the estimator falls back to a tool-level prior aggregated across programs.

\subsection{Efficient Pin with TTL in Scheduler}
After the tool call handler gives the TTL value, the scheduler will need to execute the pin operation.

\mypara{Request Pinning}
If the step is not signified to be the last step (ex. parsed to contain a tool call), the scheduler calls the tool-call handler to obtain the TTL value $\tau^{*}$ and, if not zero, invokes \texttt{pin\_request(request, $\tau^{*}$)}. 
This records a pair of request and its expiration time $\texttt{current\_timestamp} + \tau^{*}$ in a dictionary \path{pinned_requests} and deliberately skips freeing the request's KV blocks.
The \path{pinned_requests} dictionary will also be passed to the waiting queue to prioritize the scheduling of the next request in the same program.

\mypara{Request Unpinning}
At the beginning of every scheduling step, the scheduler runs \path{unpin_requests()}.
It scans \path{pinned_requests} and unpins entries whose TTL have expired \emph{and} whose \path{program_id} does not currently appear in the waiting queue.
This prevents premature eviction when a follow-up request has already arrived at the inference engine but scheduler has not been able to schedule it. 
Additionally, when a program's last step finishes, the scheduler proactively unpins any remaining pins with the same \path{program_id}, as no KV cache reuse is expected in the near future.

\mypara{Prevention of deadlocks}
Pinned requests can accumulate and potential deadlock could occur when all the GPU memory is occupied by the pinned requests. 
% When no new requests can be scheduled to execute because the space has already been occupied by the pinned requests.
% Need modification, the deadlock not explained clearly
Since the pinned requests would be preserved if the next request of the same program is still in the waiting queue, 
the entire scheduling loop could be stuck and no new requests can be scheduled to run due to the lack of space.

Thus, we need a mechanism to unpin the requests when the such a deadlock occurs.
In \name, when the scheduling logic fails to schedule a new request due to space contention, it will check if there are any pinned requests in \texttt{pinned\_requests}. If there are,
we iteratively selects victims from \texttt{pinned\_requests} with the latest program arrival time to unpin and free the space until the first request can be scheduled to run. 
The chosen request will be removed from its queue, its KV cache is freed, and it is re-queued as needed, ensuring that subsequent allocations can proceed to run. 
This prevents deadlock even when many pins are present.

\mypara{Offline Profile}
\label{sec:profile}
In order to predict the prefill time and reloading time ($\mathsf{Prefill\text{-}Reload(r)}$) based on context size as needed in Sec~\ref{sec:cost-model}, we perform an offline profile on each hardware and model pair for online estimation. We profile for two purposese: \textbf{(1)} GPU-CPU bandwidth for CPU offloading cases. We measure by taking the average CPU offloading throughput. \textbf{(2)} Prefill vs context length curve for estimating prefill cost. We measure this by doing prefill for chunk sizes $\{1000, 2000, 4000,... max\_context\_length\}$ and fit a quadratic curve on the data. Admittedly, there could be some pages for the request remaining in GPU memory that does not need recomputation. But these remaining pages are usually small when memory is contended and we approximate by the full prefill time with little error. Profiling takes less than 10 minutes for each hardware model pair.

\definecolor{colorFCFS}{HTML}{1f77b4}    % vLLM
\definecolor{colorLHC}{HTML}{2ca02c}     % Ours
\definecolor{colorPLAS}{HTML}{ff7f0e}    % Autellix
\definecolor{colorINFERCEPT}{HTML}{d62728} % InferCept

\newcommand{\legendline}[5][1.5pt]{%
  \begin{tikzpicture}[baseline=-0.6ex, scale=0.8]
    % line
    \draw[line width=1.2pt, color=#3, #5] (0,0) -- (0.9,0);
    % marker
    \node[#2, draw=#3, fill=#3, inner sep=#1, minimum size=#1*2] at (0.45,0) {};
  \end{tikzpicture}%
  \hspace{2pt}#4
}

\begin{figure*}[t]
    \centering
    
    % ---------- Legends ----------
    \begin{minipage}[t]{\columnwidth}
      \centering
      \setlength{\tabcolsep}{2pt}
      \renewcommand{\arraystretch}{1.0}
      \hspace{30pt}
      \begin{tabular}{@{}l@{\hspace{8pt}}l@{\hspace{8pt}}l@{}}
        \legendline[1.2pt]{regular polygon, regular polygon sides=4}{colorLHC}{\textbf{Ours}}{solid} &
        \legendline[1.2pt]{circle}{colorFCFS}{\textbf{vLLM}}{dashed} &
        \legendline[1pt]{regular polygon, regular polygon sides=3}{colorPLAS}{\textbf{Autellix}}{dashdotted}
      \end{tabular}
    \end{minipage}
    % \vspace{mm}
    
    % ---------- Row 1: SWE-Bench ----------
    % \begin{minipage}[t]{0.05\textwidth}
    %     \vspace{-25mm} 
    %     \raggedleft
    %     \rotatebox{90}{\small\textbf{SWE-Bench}}
    % \end{minipage}
    \hspace{2mm}
    \begin{minipage}[t]{1\textwidth}
        \centering
        \begin{minipage}[t]{0.24\textwidth}
            \centering
            \includegraphics[width=\textwidth]{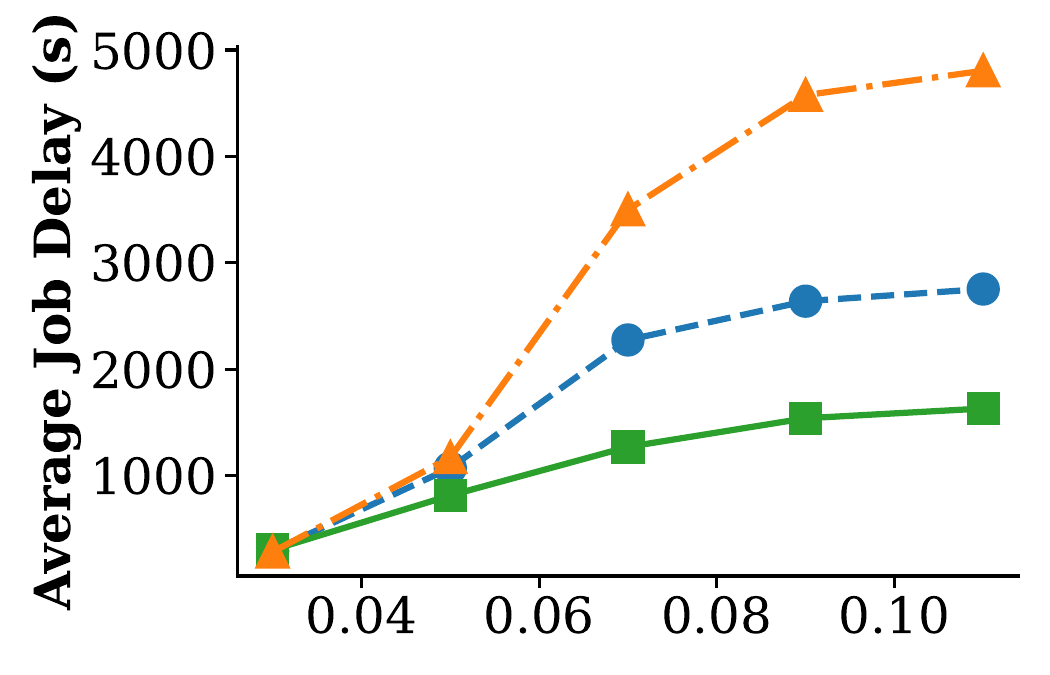}
        \end{minipage}
        \hfill
        \begin{minipage}[t]{0.24\textwidth}
            \centering
            \includegraphics[width=\textwidth]{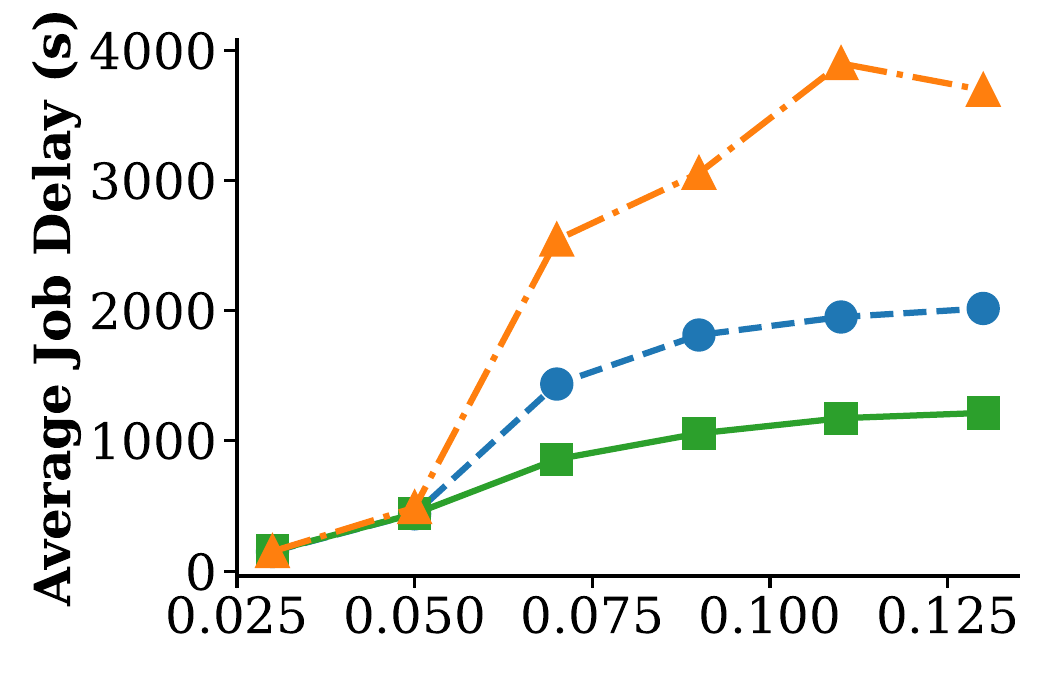}
        \end{minipage}
        \hfill
        \begin{minipage}[t]{0.24\textwidth}
            \centering
            \includegraphics[width=\textwidth]{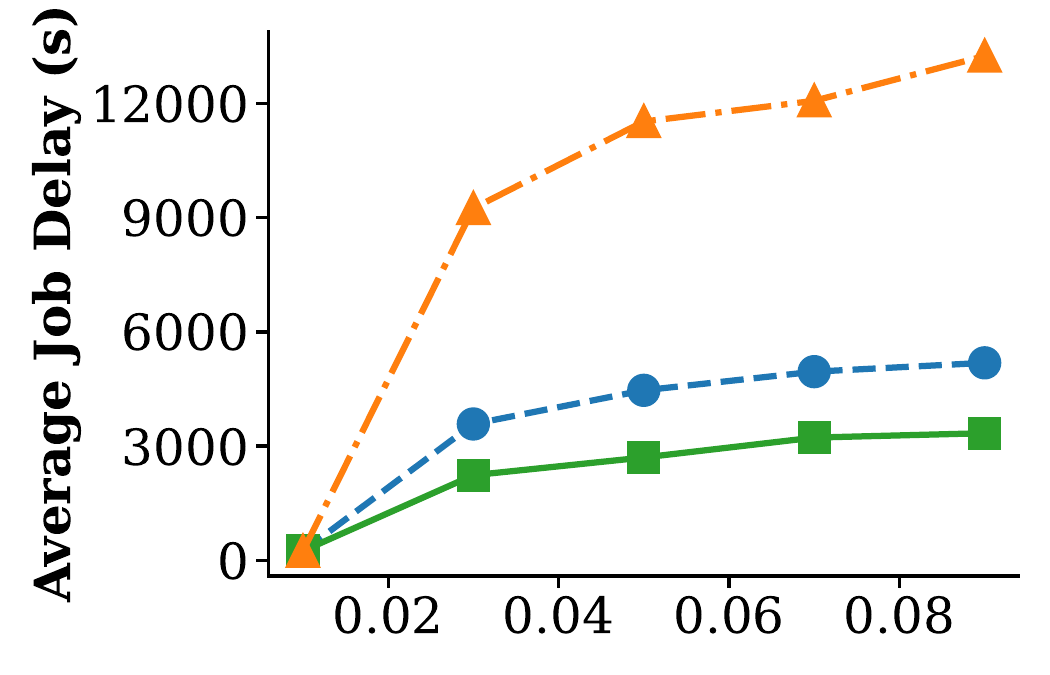}
        \end{minipage}
        \hfill
        \begin{minipage}[t]{0.24\textwidth}
            \centering
            \includegraphics[width=\textwidth]{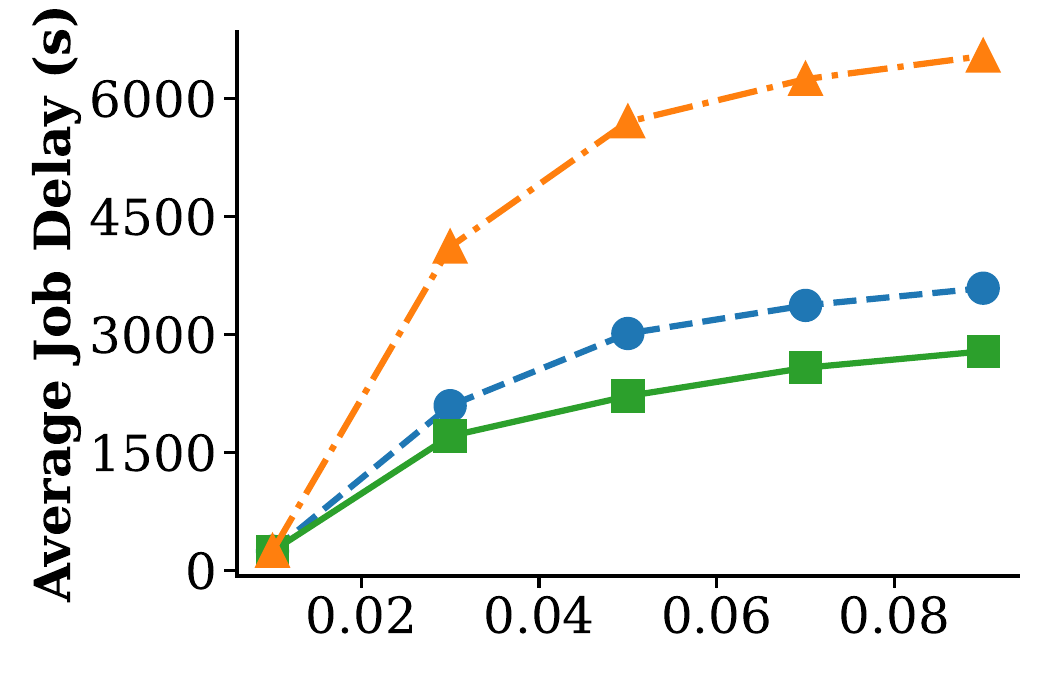}
        \end{minipage}
    \end{minipage}

    \vspace{-4pt}

    % ---------- Row 2: BFCL ----------
    % \begin{minipage}[t]{0.05\textwidth}
    %     \vspace{-21mm} 
    %     \raggedleft
    %     \rotatebox{90}{\small\textbf{BFCL}}
    % \end{minipage}
    \hspace{2mm}
    \begin{minipage}[t]{1\textwidth}
        \centering
        \begin{minipage}[t]{0.24\textwidth}
            \centering
            \includegraphics[width=\textwidth]{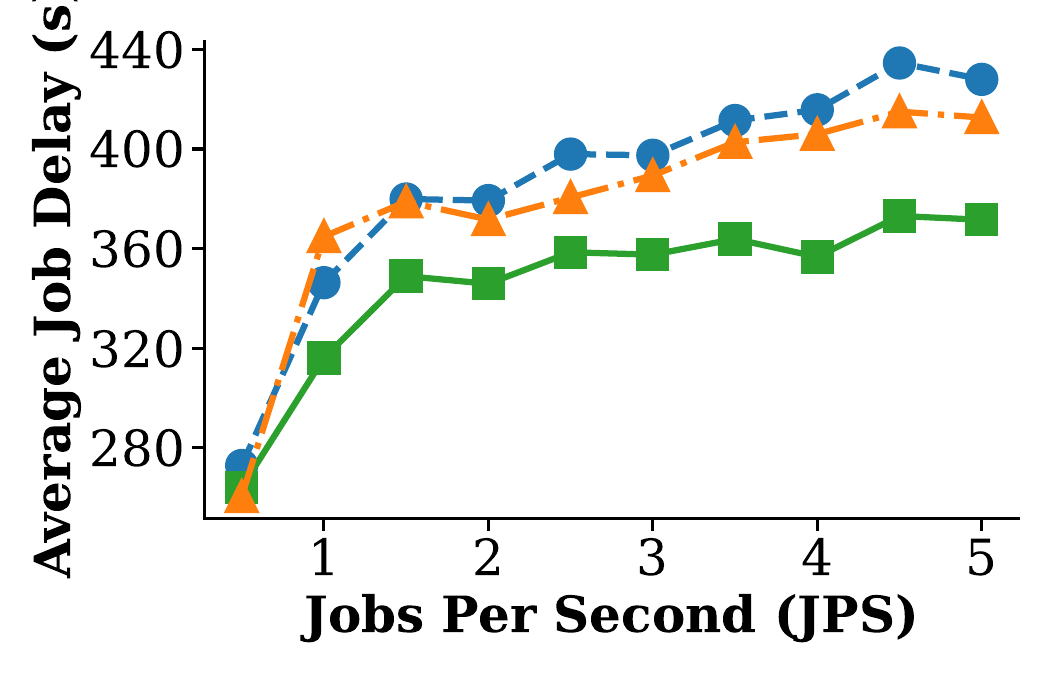}
            \vspace{2pt}
            \hspace{5mm}
            {\small \textbf{Llama 70B (4×B200)}}
        \end{minipage}
        \hfill
        \begin{minipage}[t]{0.24\textwidth}
            \centering
            \includegraphics[width=\textwidth]{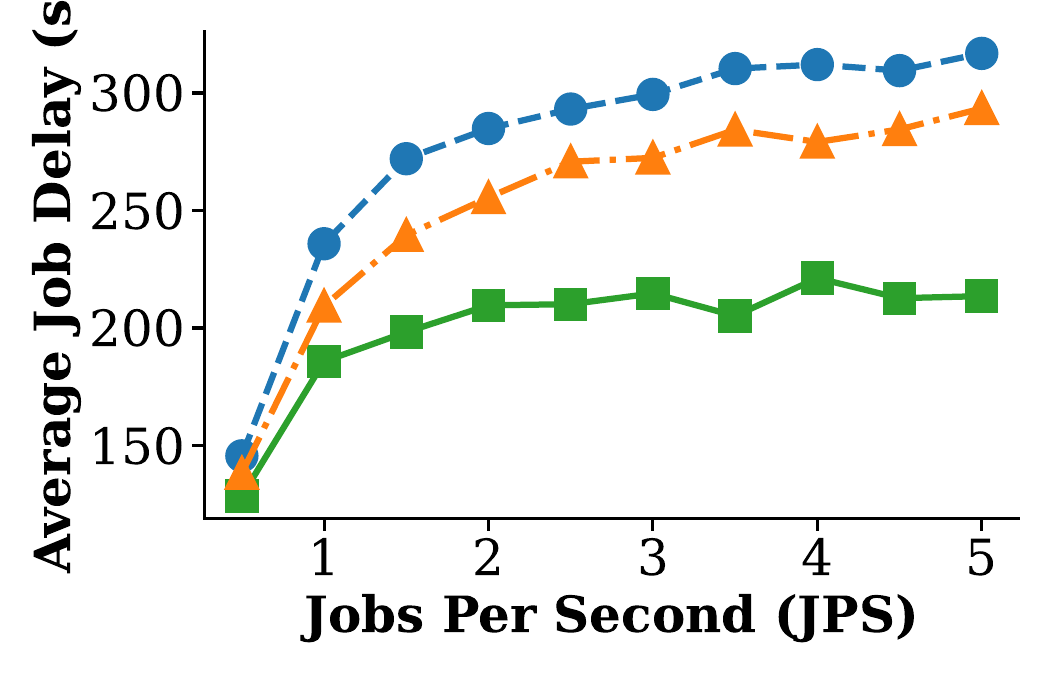}
            \vspace{2pt}
            \hspace{3mm}
            {\small \textbf{Llama 8B (1×B200)}}
        \end{minipage}
        \hfill
        \begin{minipage}[t]{0.24\textwidth}
            \centering
            \includegraphics[width=\textwidth]{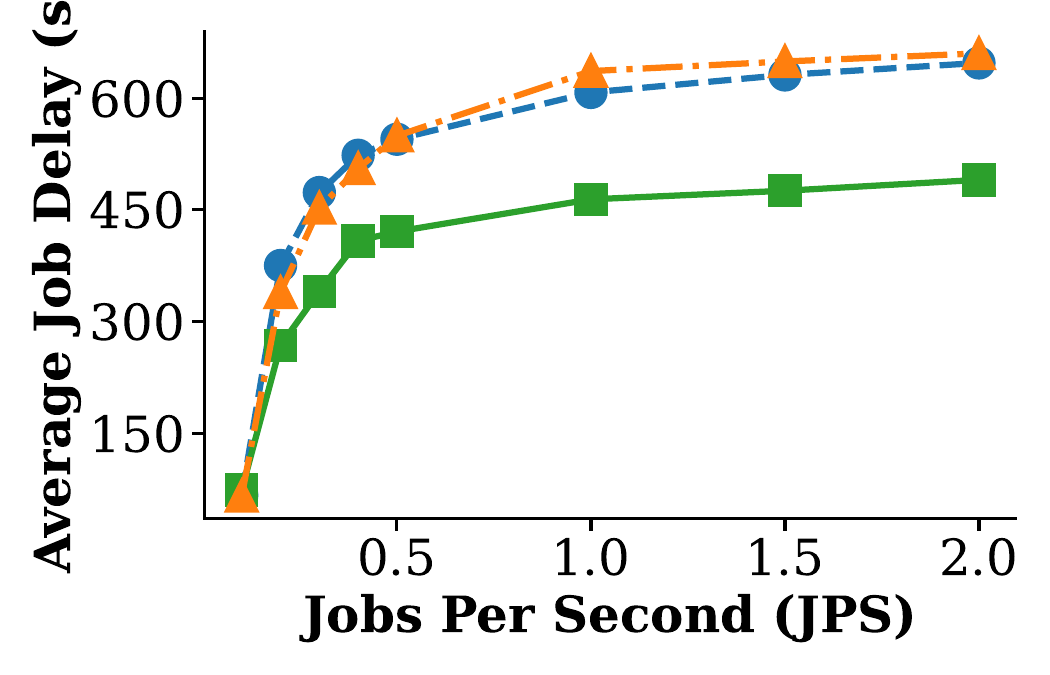}
            \vspace{2pt}
            \hspace{3.7mm}
            {\small \textbf{Llama 8B (1×A100)}}
        \end{minipage}
        \hfill
        \begin{minipage}[t]{0.24\textwidth}
            \centering
            \includegraphics[width=\textwidth]{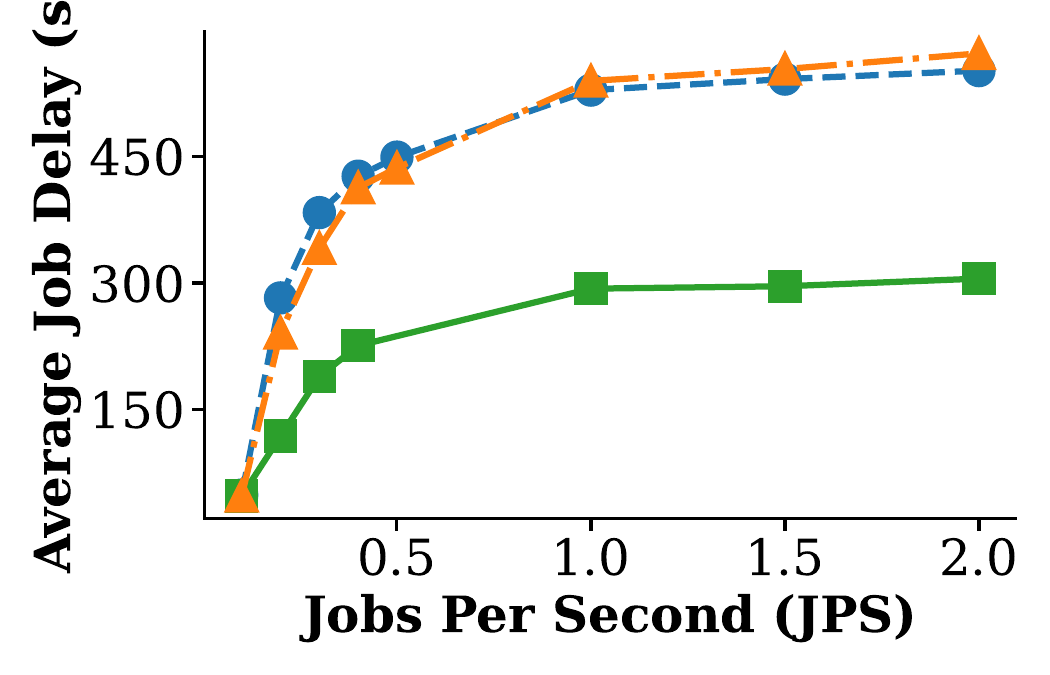}
            \vspace{2pt}
            \hspace{3mm}
            {\small \textbf{Gemma 12B (1×A100)}}
        \end{minipage}
    \end{minipage}

    \tightcaption{\name outperforms baseline schedulers across different model sizes, hardware configurations, and datasets (\textbf{Top}: SWE-bench; \textbf{Bottom}: BFCL).}
    \label{fig:eval-e2e}
\end{figure*}

\begin{figure}[t]
    \centering
    \begin{minipage}[t]{0.99\columnwidth}
      \centering
      \setlength{\tabcolsep}{2pt}
      \begin{tabular}{cccc}
      
        \legendline[1pt]{regular polygon, regular polygon sides=4}{colorLHC}{\textbf{Ours}}{solid} &
        \legendline[1pt]{circle}{colorFCFS}{\textbf{vLLM}}{dashed} &
        \legendline[1pt]{regular polygon, regular polygon sides=3}{colorPLAS}{\textbf{Autellix}}{dashdotted} &
        % \legendline[1pt]{diamond}{colorINFERCEPT}{\textbf{InferCept}}{dotted}
      \end{tabular}
    \end{minipage}
    \vspace{2mm}
    \begin{subfigure}[b]{0.23\textwidth}
        \centering
        
        \includegraphics[width=\textwidth]{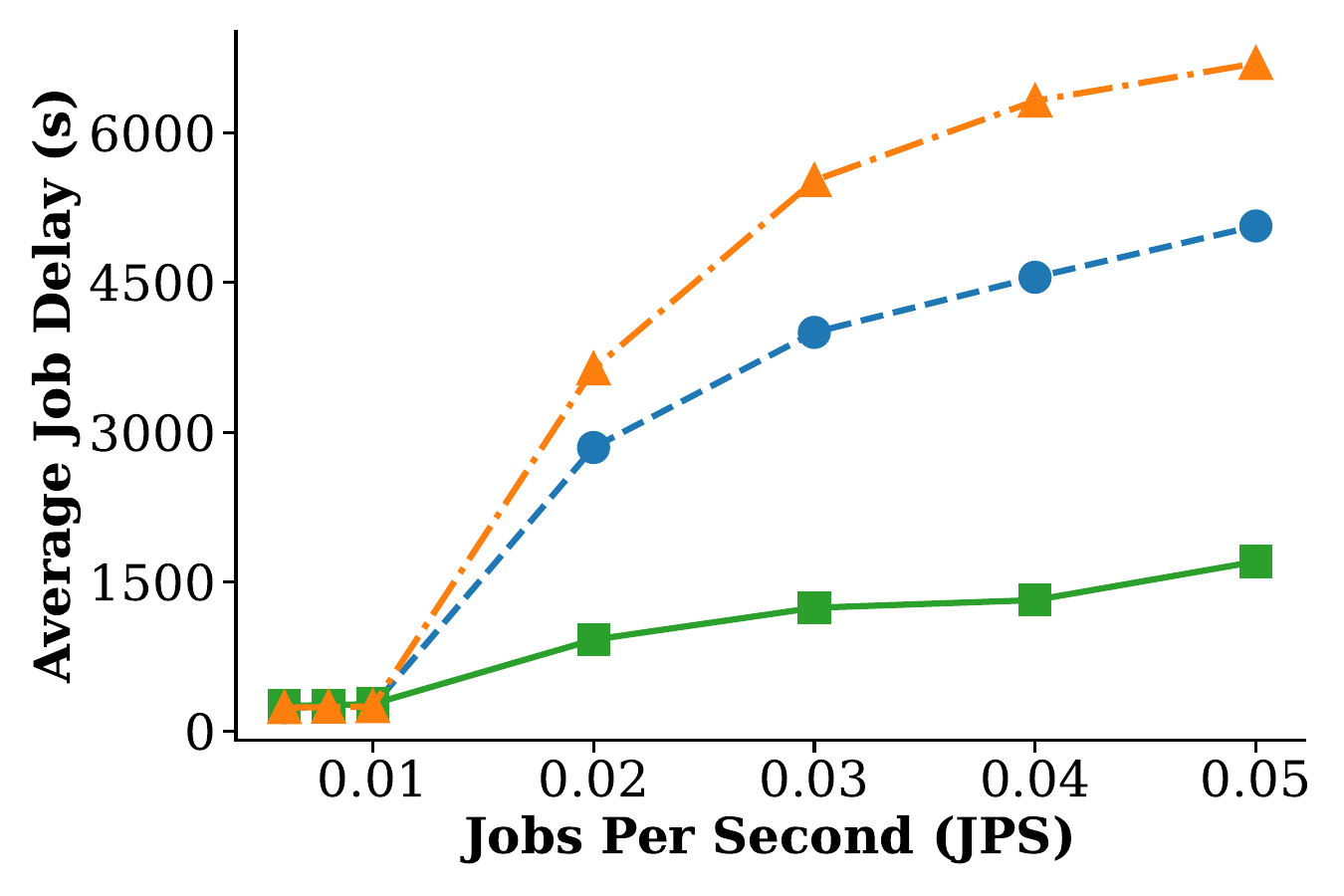}
        \tightcaption{Average}
    \end{subfigure}
    \hfill
    \begin{subfigure}[b]{0.23\textwidth}
        \centering
        \includegraphics[width=\textwidth]{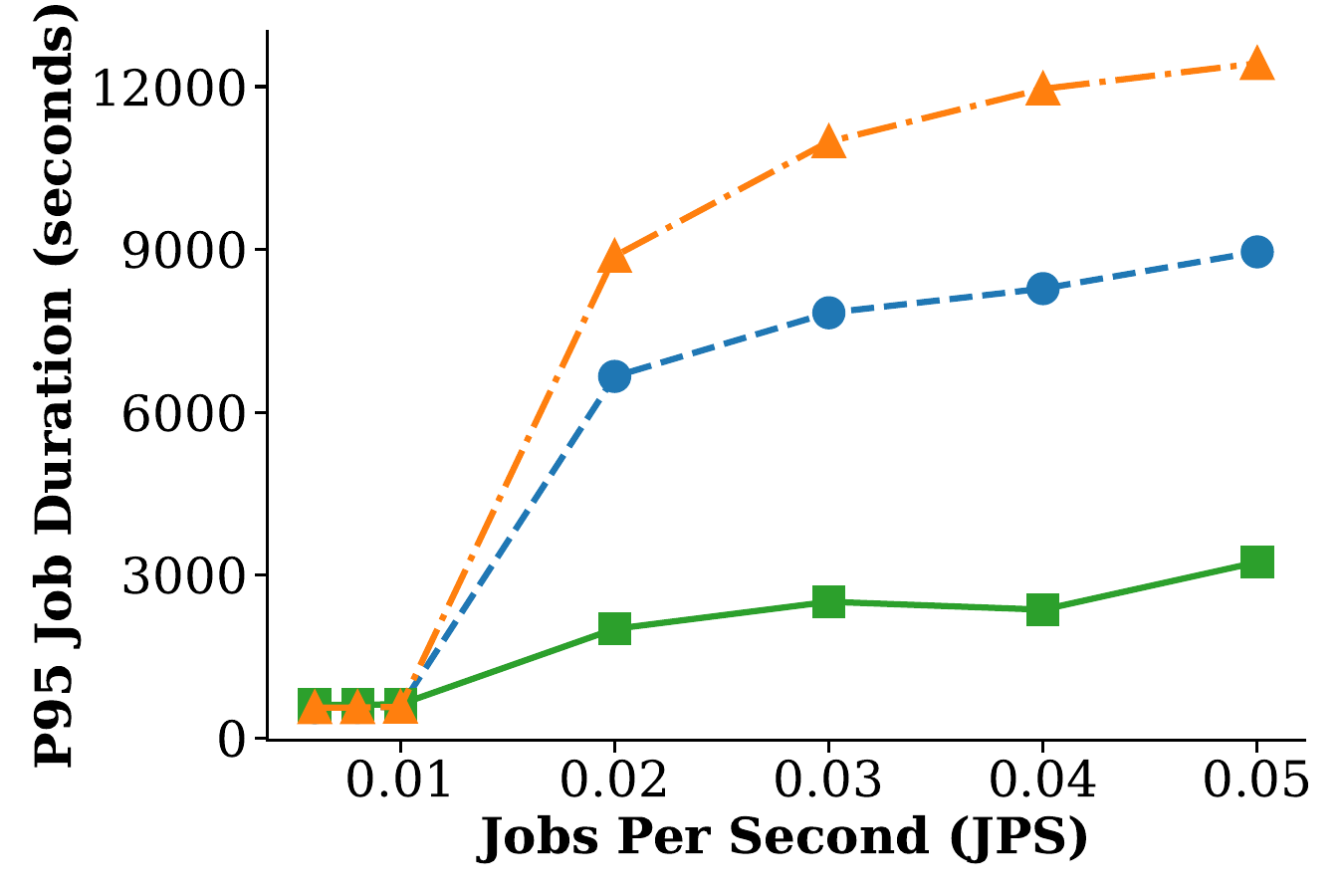}
        \tightcaption{P95}
    \end{subfigure}
    \tightcaption{\name achieves best performance on OpenHands with Llama-8B on average and P95 delays with H100.}
    \label{fig:openhands}
\end{figure}

\tightsubsection{Implementation}
We implemented \name on top of vLLM with about 1k lines of Python. Besides the above pinning operations added to the scheduler class, we use three functions from tool call handler in vLLM's original scheduler:

\begin{itemize}[leftmargin=*, itemsep=2pt]
\item \texttt{func\_call\_finish(tool, timestamp):}  When a request finishes and is parsed to contain a tool call, this function informs the tool call handler to record the tool call starting time.
\item \texttt{update\_tool\_call\_time(program\_id, timestamp):}\\ When a new request arrives, it denotes that the tool call from the previous request finished, so we record the time.
\item \texttt{set\_up\_ttl(request, tool):} Based on previous tool call information and the system setup, give the best TTL value for the scheduler for this finished request.
\end{itemize}

% \myparaq{How is this different from previous "preserve" operations}
% The main different between KV cache TTL and previous "preserve" operations such as one provided by InferCept~\cite{abhyankar2024infer} is that
% the TTL-based approach explicitly sets an expectation for the benefits of the waiting time based on the observed tool call durations based on the many-turn and variable tool call assumptions.

% Thus, \name will consider the benefits of ordering preservation during calculation of the TTL value.
% Moreover, the TTL-based approach is more robust than previous "preserve" operations. 
% By setting a TTL value for pinned requests instead of an infinite preservation, \name can automatically release memory from requests that exceed their expected tool call duration, preventing performance degradation under variable tool call itme.

\begin{figure*}[h]
    \centering
    % ---------- Legend ----------
    \begin{minipage}[t]{\textwidth}
      \centering
      \setlength{\tabcolsep}{4pt}
      \hspace{30pt}
      \begin{tabular}{cccc}
        \legendline[1.2pt]{regular polygon, regular polygon sides=4}{colorLHC}{\textbf{Ours}}{solid} &
        \legendline[1.2pt]{circle}{colorFCFS}{\textbf{vLLM}}{dashed} &
        \legendline[1pt]{regular polygon, regular polygon sides=3}{colorPLAS}{\textbf{Autellix+}}{dashdotted} &
        \legendline[1.2pt]{diamond}{colorINFERCEPT}{\textbf{InferCept}}{dotted}
      \end{tabular}
    \end{minipage}

    % ---------- Row 1: SWE-Bench ----------
    % Dataset labels are described in the caption to keep all panels full width.
    \hspace{2mm}
    \begin{minipage}[t]{1\textwidth}
        \centering
        \begin{minipage}[t]{0.24\textwidth}
            \centering
            \includegraphics[width=\textwidth]{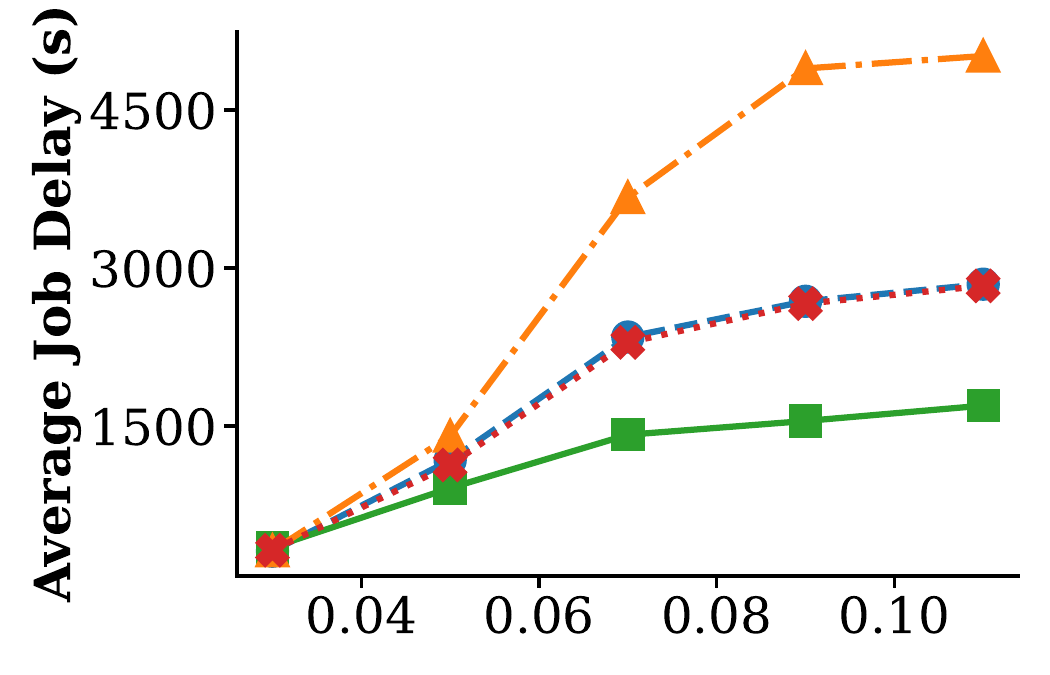}
        \end{minipage}
        \hfill
        \begin{minipage}[t]{0.24\textwidth}
            \centering
            \includegraphics[width=\textwidth]{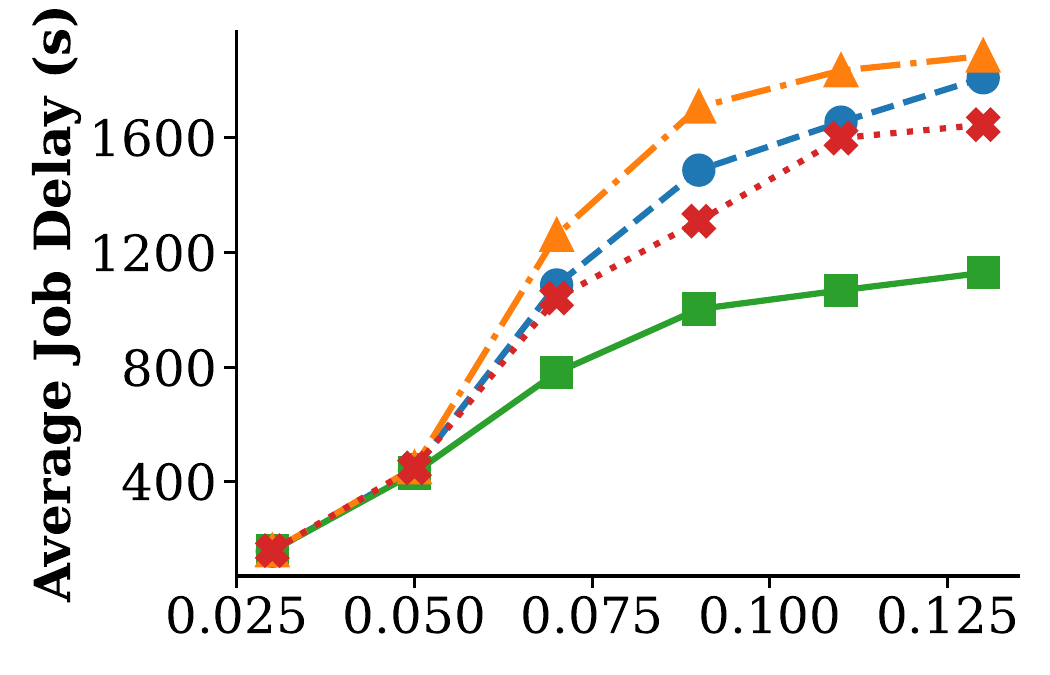}
        \end{minipage}
        \hfill
        \begin{minipage}[t]{0.24\textwidth}
            \centering
            \includegraphics[width=\textwidth]{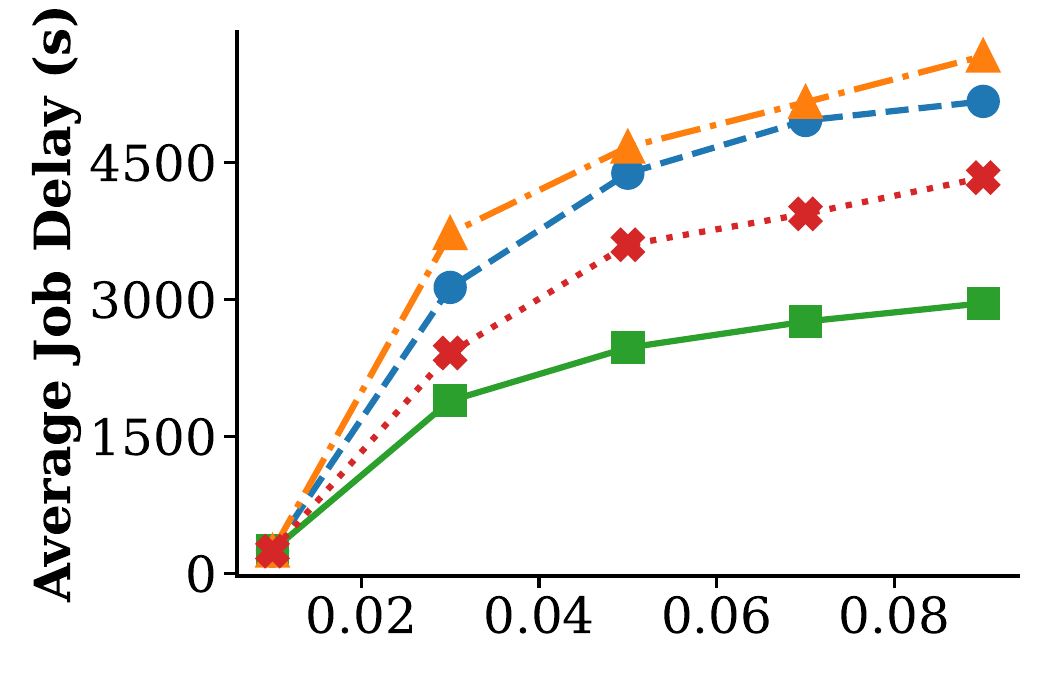}
        \end{minipage}
        \hfill
        \begin{minipage}[t]{0.24\textwidth}
            \centering
            \includegraphics[width=\textwidth]{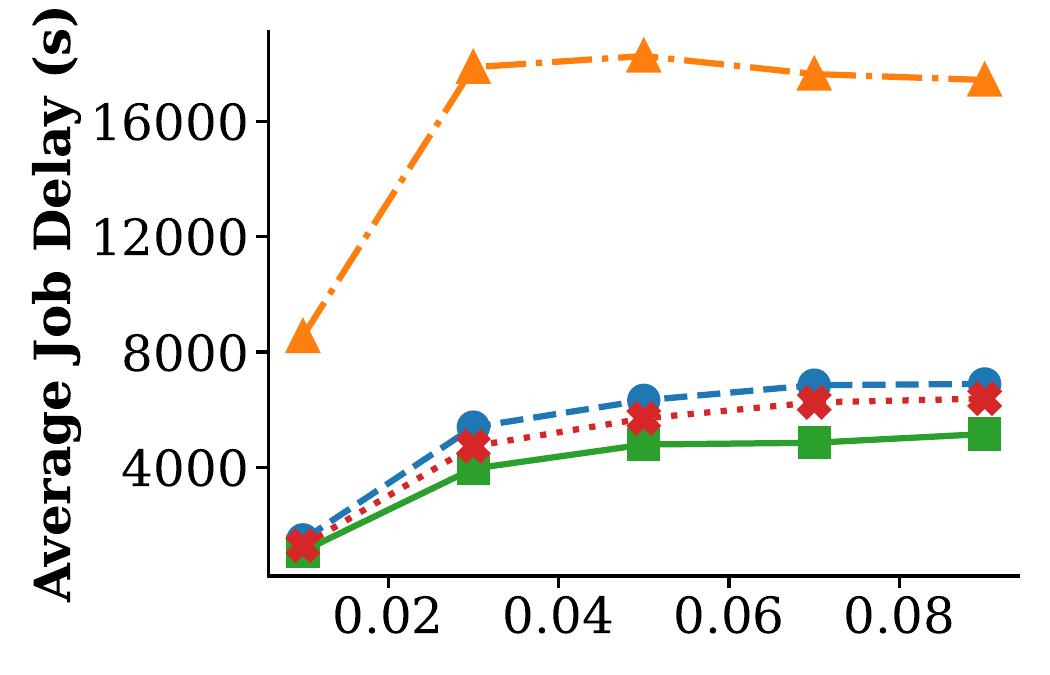}
        \end{minipage}
    \end{minipage}

    \vspace{-4pt}

    % ---------- Row 2: BFCL ----------
    \hspace{2mm}
    \begin{minipage}[t]{1\textwidth}
        \centering
        \begin{minipage}[t]{0.24\textwidth}
            \centering
            \includegraphics[width=\textwidth]{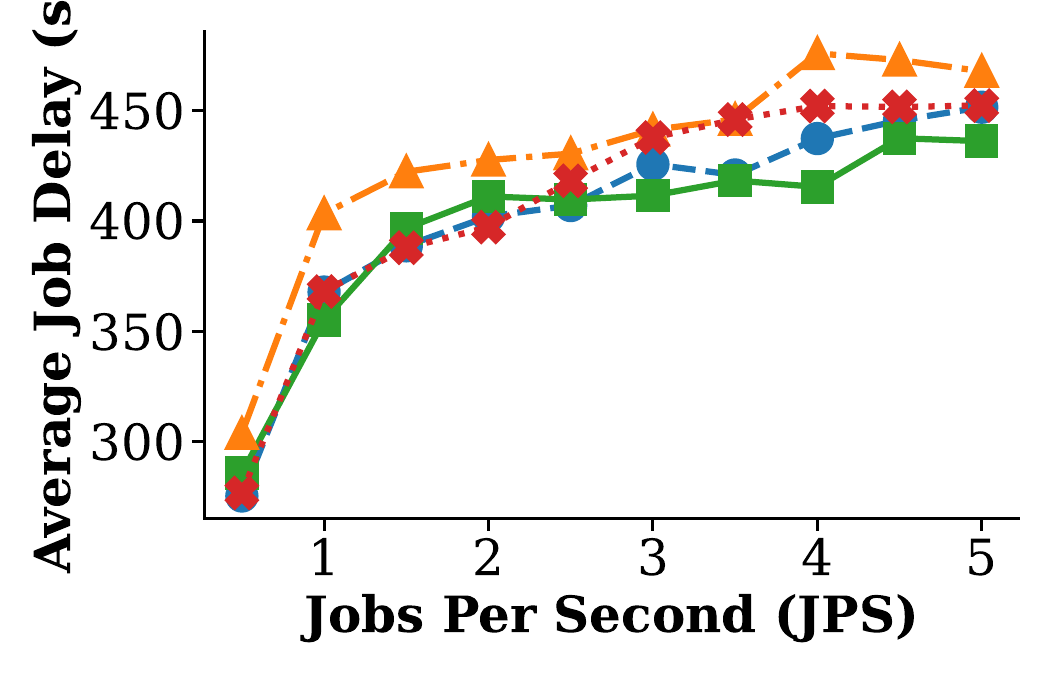}
            \vspace{2pt}
            {\small\textbf{Llama 70B (4$\times$B200)}}
        \end{minipage}
        \hfill
        \begin{minipage}[t]{0.24\textwidth}
            \centering
            \includegraphics[width=\textwidth]{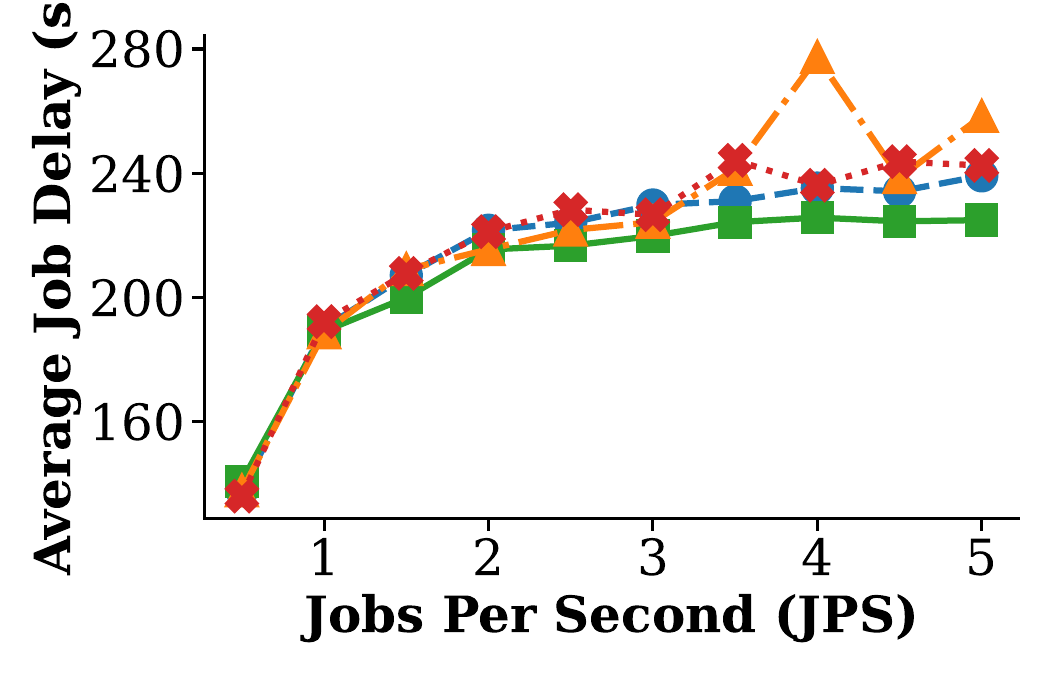}
            \vspace{2pt}
            {\small\textbf{Llama 8B (1$\times$B200)}}
        \end{minipage}
        \hfill
        \begin{minipage}[t]{0.24\textwidth}
            \centering
            \includegraphics[width=\textwidth]{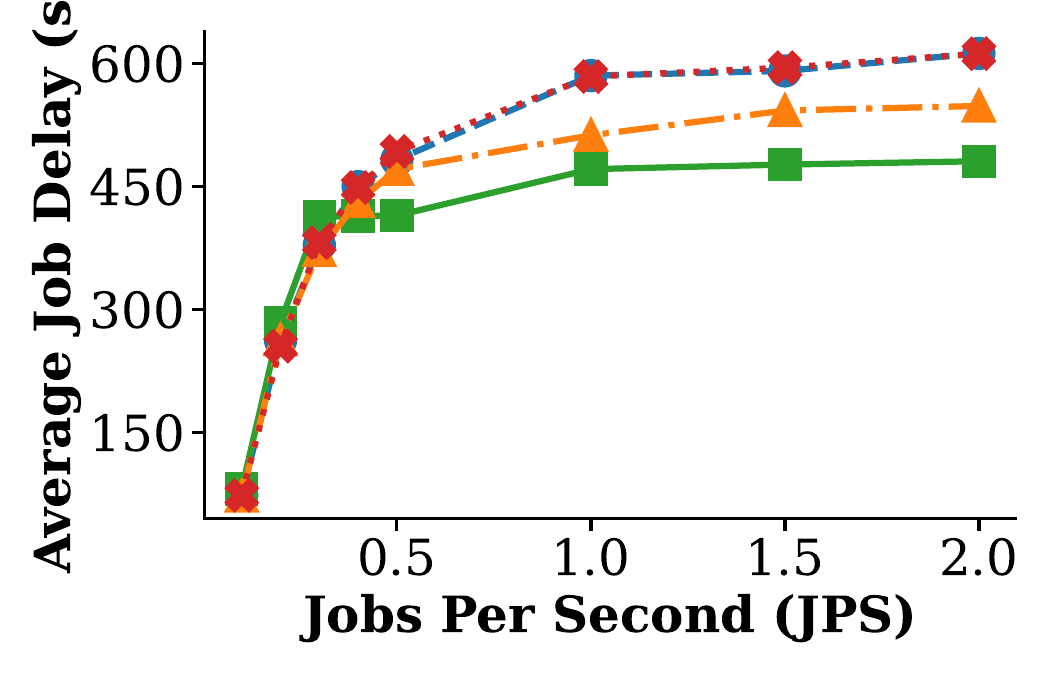}
            \vspace{2pt}
            {\small\textbf{Llama 8B (1$\times$A100)}}
        \end{minipage}
        \hfill
        \begin{minipage}[t]{0.24\textwidth}
            \centering
            \includegraphics[width=\textwidth]{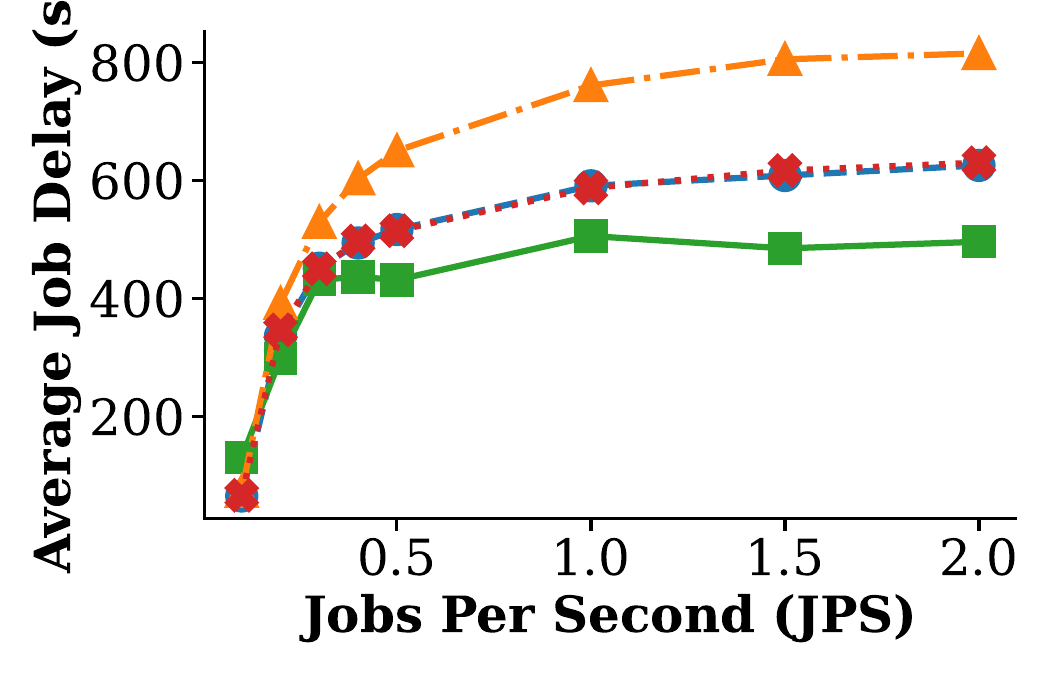}
            \vspace{2pt}
            {\small\textbf{Gemma 12B (1$\times$A100)}}
        \end{minipage}
    \end{minipage}

    \tightcaption{\name achieves consistent improvement when DRAM offloading is enabled. It improves over systems with smart DRAM offloading logic like InferCept by considering tool-call and multi-turn together (\textbf{Top}: SWE-Bench; \textbf{Bottom}: BFCL).}
    \label{fig:eval-e2e-cpu}
\end{figure*}

\tightsection{Evaluation}
% \lhc{ablation}
% \lhc{add error bar}
\label{sec:evaluation}
Our key takeaways from the evaluation are:
\begin{itemize}[leftmargin=*]
    \item \mypara{Delay Reduction} \name achieves significant delay reduction improvements over baseline schedulers through intelligent KV cache pinning
    \item \mypara{Robustness} \name outperforms baselines across turn counts and different offloading scenarios.
    \item \mypara {Out-of-the-Box Usability} \name can be used to run real agent faster without a quality drop.
\end{itemize}

\tightsubsection{Setup}
\mypara{Model and Hardware} 
We evaluate \name with Llama-3.1-8B, Llama-3.1-70B, and Gemma-3-12B. 
We use A100-SXM GPU from Runpod, H100 from AWS and \company, and B200 GPU from on-prem servers. 

\mypara{Datasets}
For results other than the real SWE-Bench experiments in Figure~\ref{fig:real_exp}, we evaluate on two collected workloads running GPT-5~\footnote{We use GPT-5 for the better model capabilities to ensure that the workflow generated are mostly correct. Base small models often fail to accomplish the task} and using poisson distribution for the arrival pattern of agent programs: 

% \lhc{runyuan, give the details of the datasets}
\begin{packeditemize}
    \item SWE-Bench~\cite{jimenez2023swe}: We run mini-swe-agent~\cite{lieret2025miniSWEagent}~\footnote{SWE-bench official agent that rank \#5 on leaderboard by Apr 13th} on SWE-Bench. We keep requests within the context window.
    \item Berkeley Function Calling Leaderboard~\cite{mao2025bfclv4web}: We used the latest version of BFCL V4 (Web Search category). This includes agents answering questions with web browsing tools. We scaled down the workload by 0.4 to fit at least 100 request in the context window of llama-3.1 (128k tokens).
    \item OpenHand~\cite{wang2025openhandsopenplatformai}: OpenHands is a popular open-source coding agent. We run the multi-SWE-bench~\cite{zan2025multiswebenchmultilingualbenchmarkissue} example in the official repo for the Go language.
\end{packeditemize}

\mypara{Main Baselines}
\begin{packeditemize}
    \item \textit{Vanilla vLLM} We use the stable release of vllm 0.10.2 with default setting, where chunk size is enabled with size 2048. 
    \item \textit{CPU DRAM offloading} We use vllm 0.10.2 with LMCache 0.3.7~\cite{cheng2025lmcache}. For A100 GPUs, we set the DRAM size used in offloading to be 100GB; For B200 and H100 GPUs, we set the DRAM size used in offloading to be 200GB per GPU. We also apply this on top of algorithms below. 
    \item \textit{Autellix} We implemented the algorithm of PLAS from Autellix~\cite{luo2025autellix} on top of vllm. We extend Autellix to CPU offloading cases by enabling LMCache (Autellix+).
    \item \textit{InferCept} We implemented the selectively preserve, swap, or evict algorithm of 
    InferCept~\cite{abhyankar2024infer} on top of vllm + lmcache. 
    Since the CPU offloading in LMCache is non-blocking (better than original InferCept), we update the cost estimation accordingly. 
    \item \textit{Distributed Inference} For real agent experiments, we compare with other open-source solutions including SGLang \path{0.5.5.post3}~\cite{sglang2025} with native cache-aware routing and Nvidia Dynamo \path{0.7.0.post1}~\cite{dynamo2025} configured with 1P1D for PD disaggregation.
    % We show results of original InferCept code in section \ref{sec:ablation}.
\end{packeditemize}

\begin{figure}[t]
    \centering
    \begin{minipage}[t]{\columnwidth}
      \centering
      \setlength{\tabcolsep}{2pt}
      \begin{tabular}{cccc}
      
        \legendline[1pt]{regular polygon, regular polygon sides=4}{colorLHC}{\textbf{Ours}}{solid} &
        \legendline[1pt]{circle}{colorFCFS}{\textbf{vLLM}}{dashed} &
        \legendline[1pt]{regular polygon, regular polygon sides=3}{colorPLAS}{\textbf{Autellix+}}{dashdotted} &
        \legendline[1pt]{diamond}{colorINFERCEPT}{\textbf{InferCept}}{dotted}
      \end{tabular}
    \end{minipage}
    \vspace{2mm}
    \begin{subfigure}[b]{0.23\textwidth}
        \centering
        
        \includegraphics[width=\textwidth]{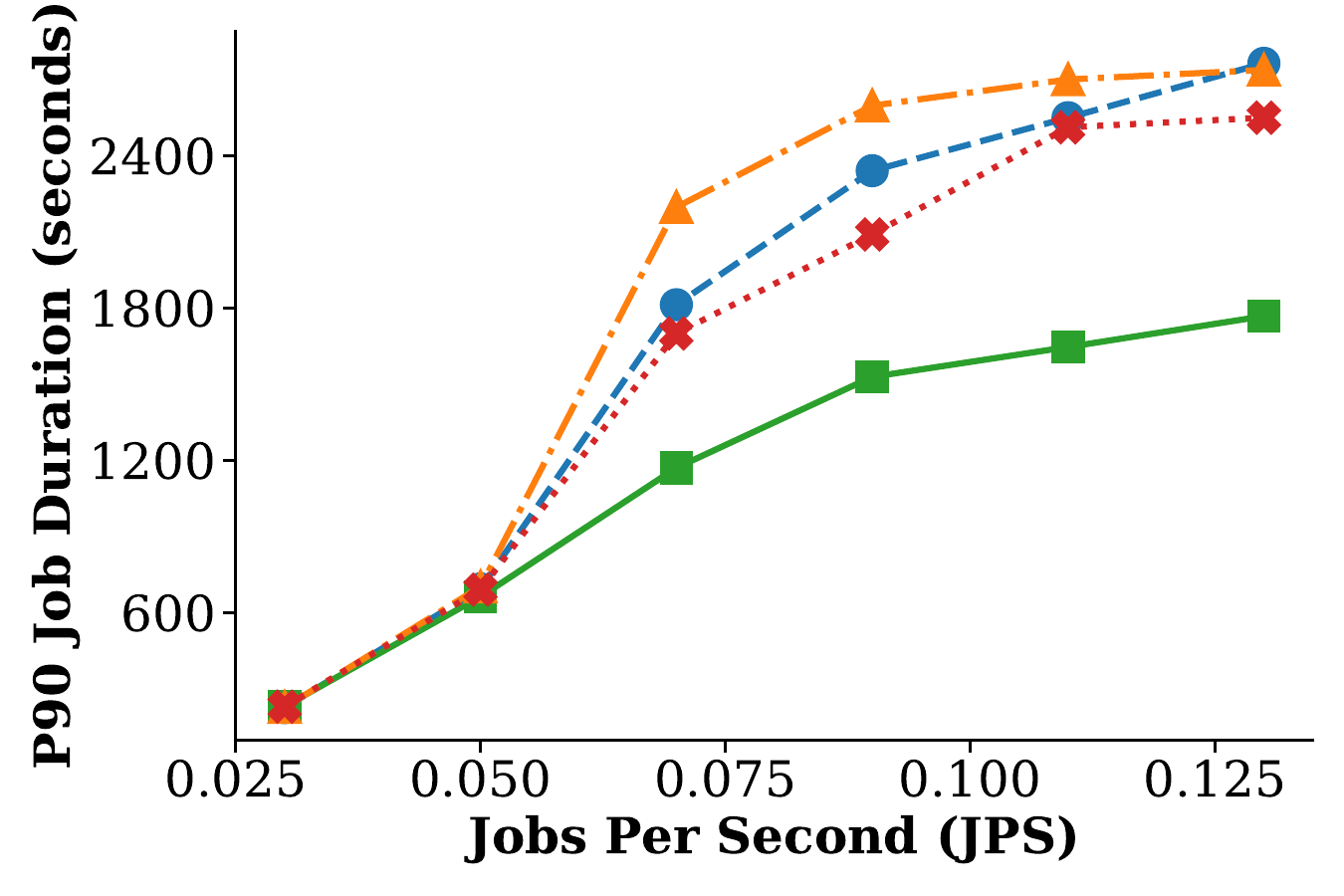}
        \tightcaption{p90}
    \end{subfigure}
    \hfill
    \begin{subfigure}[b]{0.23\textwidth}
        \centering
        \includegraphics[width=\textwidth]{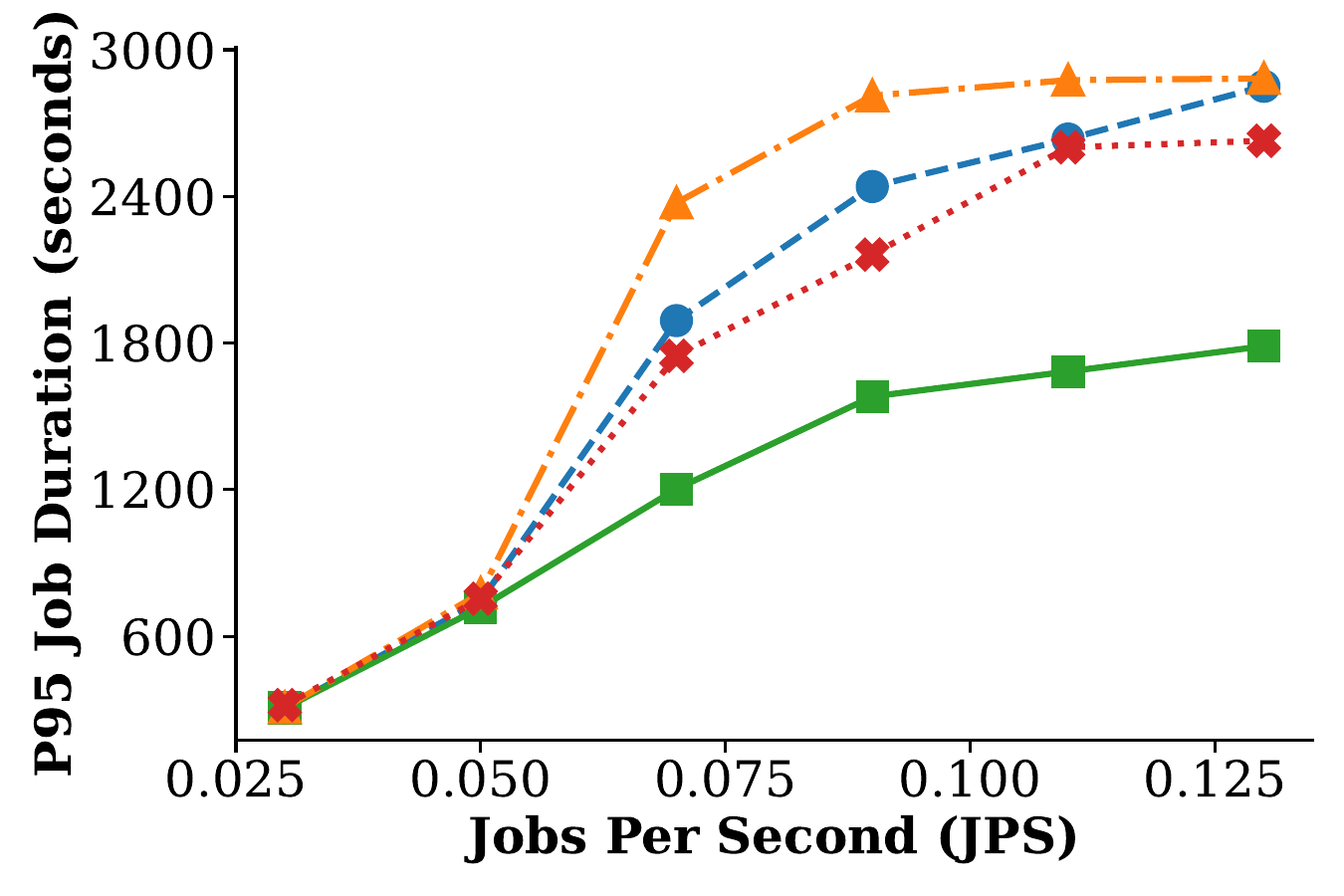}
        \tightcaption{p95}
    \end{subfigure}
    \tightcaption{\name achieves better P90 and P95 latency for running SWE Bench trace with Llama-8B model.}
    \label{fig:p95}
\end{figure}

% \vspace{-4pt}
\definecolor{colorSGLang}{HTML}{9B59B6}

% Dynamo
\definecolor{colorDynamo}{HTML}{17becf}
\begin{figure}[t]
    \centering
    \begin{minipage}[t]{0.99\columnwidth}
  \centering
  \setlength{\tabcolsep}{4pt}
  \begin{tabular}{ccc}
    \legendline[1pt]{regular polygon, regular polygon sides=4}{colorLHC}{\textbf{Ours}}{solid} &
    \legendline[1pt]{regular polygon, regular polygon sides=3}{colorSGLang}{\textbf{SGLang}}{dashdotted} &
    \legendline[1pt]{diamond}{colorDynamo}{\textbf{Dynamo}}{dotted}
  \end{tabular}
\end{minipage}

    \begin{subfigure}[b]{0.23\textwidth}
        \centering
        \includegraphics[width=\textwidth]{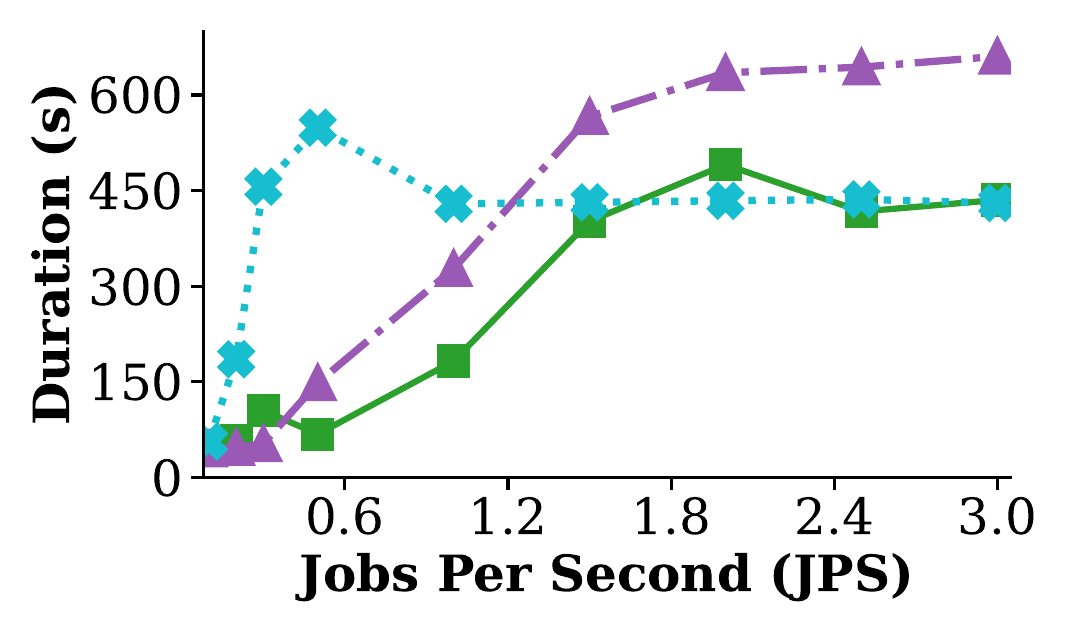}
    \end{subfigure}
    \hfill
    \begin{subfigure}[b]{0.23\textwidth}
        \centering
        \includegraphics[width=\textwidth]{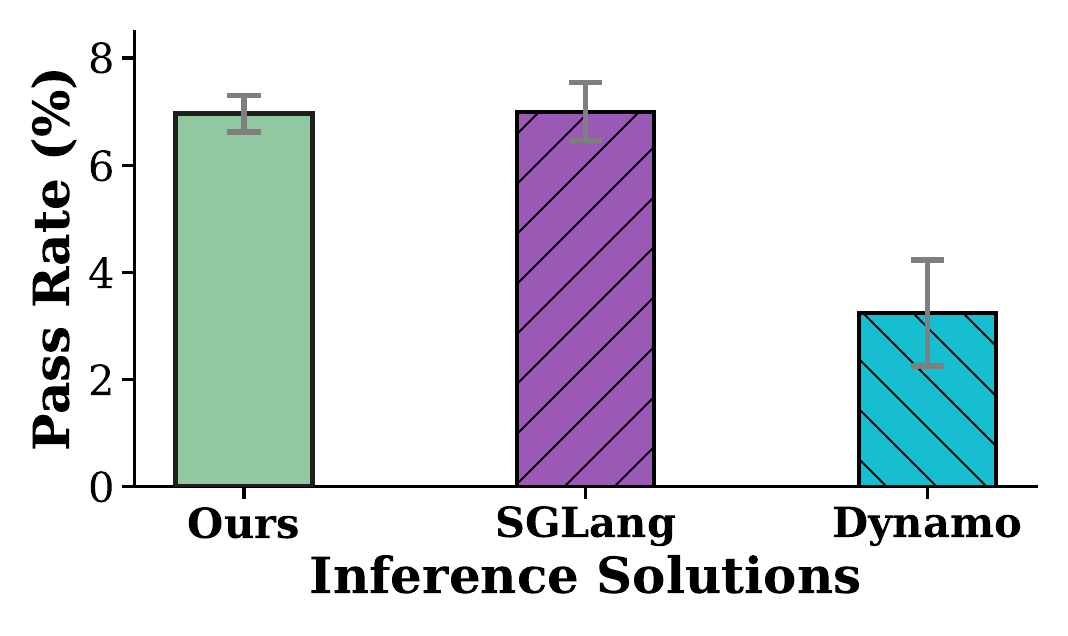}
    \end{subfigure}
    \tightcaption{\name improves delay under the pass rate for real SWE-agents in distributed settings.}
    \label{fig:real_exp}

\end{figure}
\tightsubsection{End-to-End Experiments}
% \lhc{need to say trace replay experiment, llama is only used to extract timing}
%\lhc{need to add InferCept}
We conduct the trace replay experiments for SWE-Bench, BFCL, and OpenHands workloads.
Figure~\ref{fig:eval-e2e}, Figure~\ref{fig:eval-e2e-cpu}, and Figure~\ref{fig:openhands} demonstrate the end-to-end improvement of \name. 
We show significant improvements in both average response time and throughput across both the BFCL and SWE-Bench workloads. 
For instance, with the Llama-3.1-8B model, \name achieves up to a 2x reduction in average response time compared to the vanilla vLLM baseline.
The performance gains are consistent across different model sizes and hardware configurations, demonstrating the effectiveness of our approach in diverse scenarios. 
Although Autellix outperforms baselines in BFCL, it underperforms in SWE-Bench due to its false assumption that requests have longer expected finish time if they execute for longer.
Note that the job per second rates are less than job per second reported in previous LLM serving papers. 
This is because agentic workloads are much more complex and can often involve more than 10 LLM inferences requests, incurring higher computational load.

We also extended our evaluation to other practical agents. As demonstrated in Figure~\ref{fig:openhands}, we achieve better delay running OpenHands agent with Llama 8B on one H100 GPU from AWS. Since the average turn number count is higher, our improvement is even more significant due to the deterioration of baselines under high turn numbers.

Moreover, we observe that \name consistently outperforms CPU offloading baselines. On the other hand, PLAS's gain on CPU offloading diminished compared with baseline. 
This demonstrates \name's robust performance improvement on scheduling bubble reduction that is orthogonal to DRAM offloading techniques.

In Figure~\ref{fig:p95}, we show that \name achieves better P90 and P95 latency due to its ability to reduce the per-turn queueing delay compared with baselines. The setup for each individual point is running Llama-8B model with a single B200 with CPU offloading set as 200GB per GPU.

% The improvements are primarily driven by our intelligent KV cache pinning strategy, which minimizes costly re-prefills and reduces scheduling bubbles, as discussed in ~\cref{sec:motivation-pin}.
\begin{figure*}[t]
    \centering
    \includegraphics[width=0.96\linewidth]{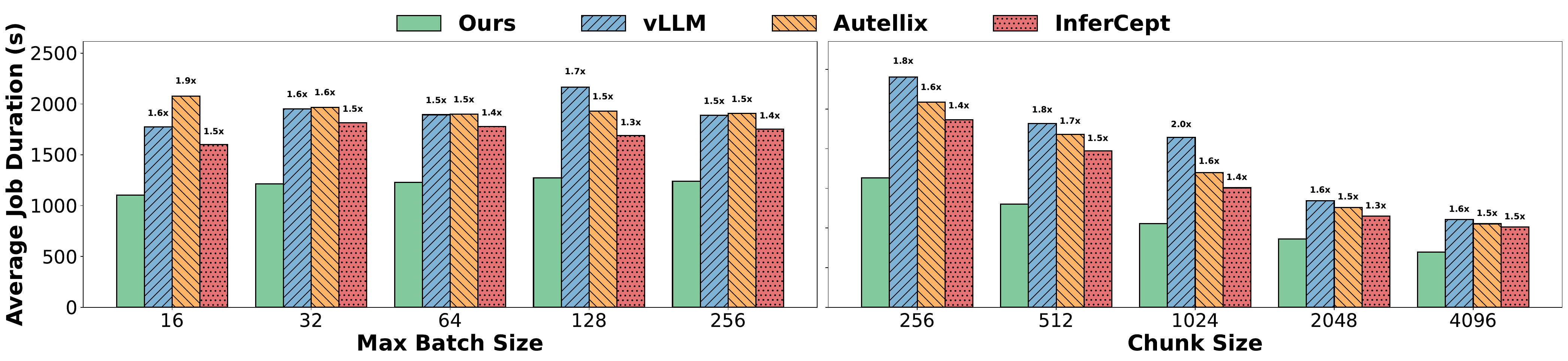}

    \tightcaption{\name improves delay across different max batch-size and chunk-size configurations.}
    \label{fig:eval_batch_chunk}
\end{figure*}

% \subsection{Improvement Analysis}
% bubble reduction graph over time
% \hry{LHC: See figure ~\ref{fig:waiting_time_analysis_comparison}}
% \lhc{can we get a graph similar to this for different workloads}

% \lhc{need to add InferCept ablation for original infercept (lets save it for later)}

% In the experiment below, we set the fixed pin timeout to be the mean tool call duration for each workload.
\begin{figure}[t]
    \centering
    \includegraphics[width=\linewidth]{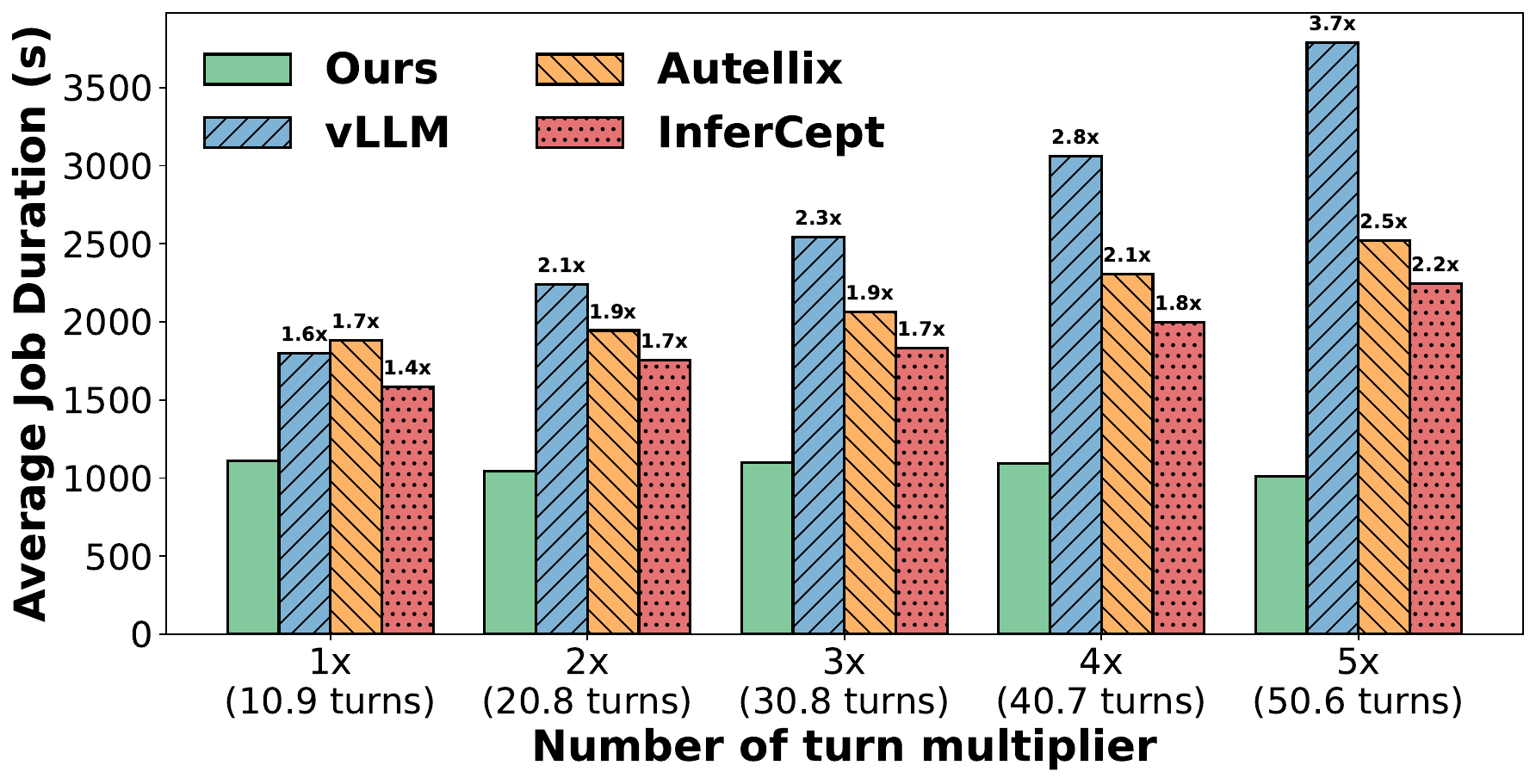}
    
    \tightcaption{\name shows greater improvement as the number of turns increases, while the delay time remains stable.}
    \label{fig:robust_study}
\end{figure}

\begin{figure}[t]
    \centering
    \begin{minipage}[t]{\columnwidth}
      \centering
      \setlength{\tabcolsep}{2pt}
      \begin{tabular}{cccc}
      
        \legendline[1pt]{regular polygon, regular polygon sides=4}{colorLHC}{\textbf{Ours}}{solid} &
        \legendline[1pt]{circle}{colorFCFS}{\textbf{vLLM}}{dashed} &
        \legendline[1pt]{regular polygon, regular polygon sides=3}{colorPLAS}{\textbf{Autellix+}}{dashdotted} &
        \legendline[1pt]{diamond}{colorINFERCEPT}{\textbf{InferCept}}{dotted}
      \end{tabular}
    \end{minipage}
    \begin{subfigure}[b]{0.23\textwidth}
        \centering
        \includegraphics[width=\textwidth]{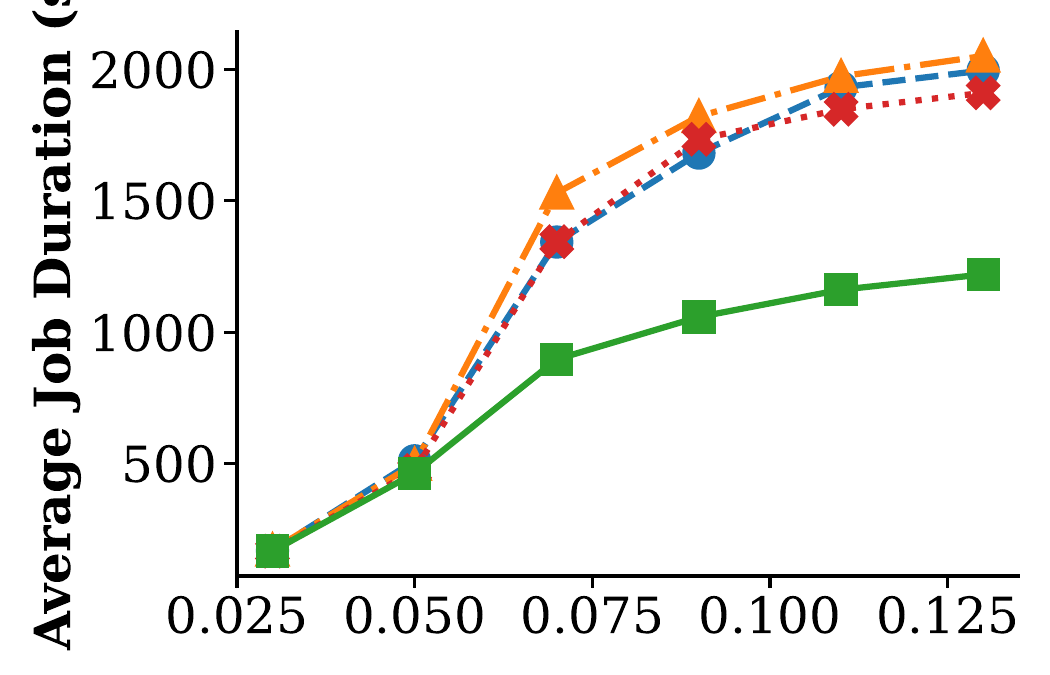}
        \subcaption{SSD Size = 400G}
        % \tightcaption{}
    \end{subfigure}
    \hfill
    \begin{subfigure}[b]{0.23\textwidth}
        \centering
        \includegraphics[width=\textwidth]{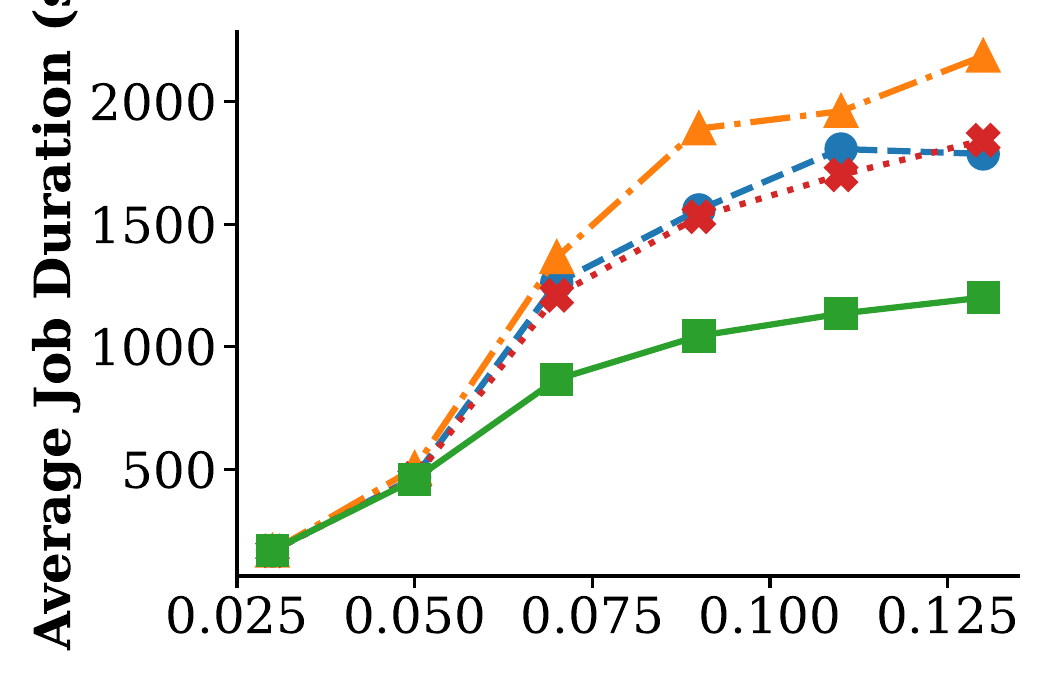}
        \subcaption{SSD Size = 800G}
        % \tightcaption{}
    \end{subfigure}
    \tightcaption{\name reduces delay when we extend offloading devices to SSDs beyond CPU offloading.}
    \label{fig:ssd}

\end{figure}

% \hry{LHC: Use figure ~\ref{fig:pin_ablation}}
\definecolor{colorNOPIN}{HTML}{6f6e6a}   % Program Level FCFS
\definecolor{colorMAGICPIN}{HTML}{8856a7} % Simplified Version

\begin{figure}[t]
    \centering
\begin{minipage}[t]{\columnwidth}
  \centering
  \setlength{\tabcolsep}{1pt}
  \begin{tabular}{cc}
    \tikz[baseline=(B.base)]{\node[inner sep=1pt] (B) {\legendline[1pt]{regular polygon, regular polygon sides=4}{colorLHC}{\textbf{Ours}}{solid}};} &
    \tikz[baseline=(B.base)]{\node[inner sep=1pt] (B) {\legendline[1pt]{circle}{colorFCFS}{\textbf{vLLM}}{dashed}};} \\
    \tikz[baseline=(B.base)]{\node[inner sep=1pt] (B) {\legendline[1pt]{diamond}{colorNOPIN}{\textbf{Program FCFS}}{dotted}};} &
    \tikz[baseline=(B.base)]{\node[inner sep=1pt] (B) {\legendline[1pt]{regular polygon, regular polygon sides=3}{colorMAGICPIN}{\textbf{Static TTL}}{dashdotted}};}
  \end{tabular}
  \vspace{2ex}
\end{minipage}
    % vspace{2mm}
    \begin{subfigure}[b]{0.23\textwidth}
        \centering
        \includegraphics[width=\textwidth]{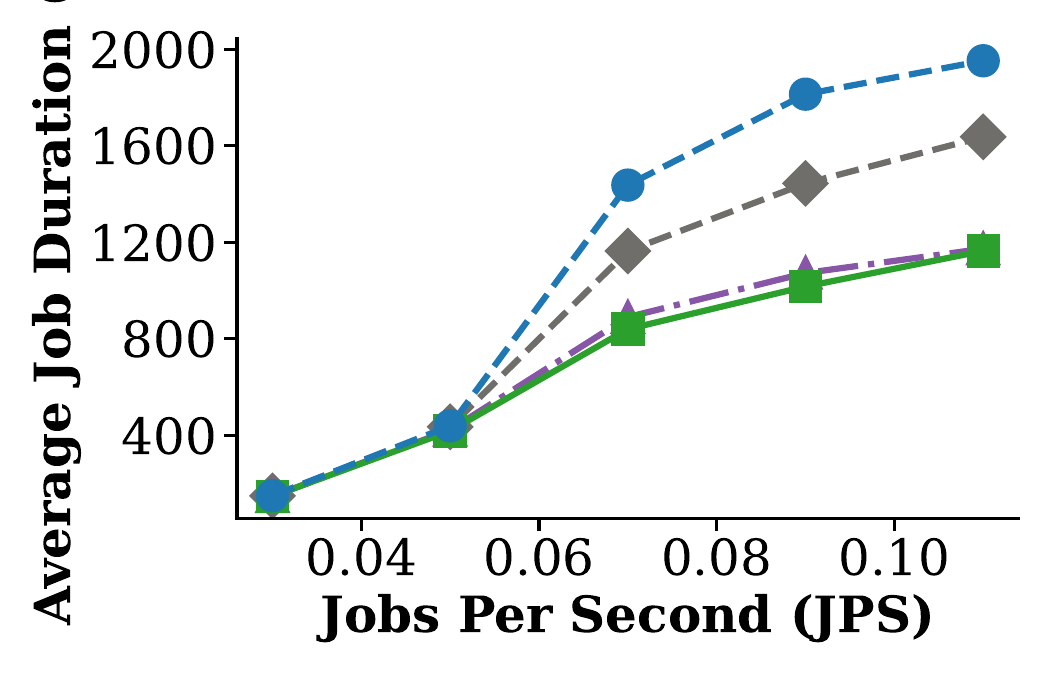}
        \subcaption{SWE-Bench}
    \end{subfigure}
    \hfill
    \begin{subfigure}[b]{0.23\textwidth}
        \centering
        \includegraphics[width=\textwidth]{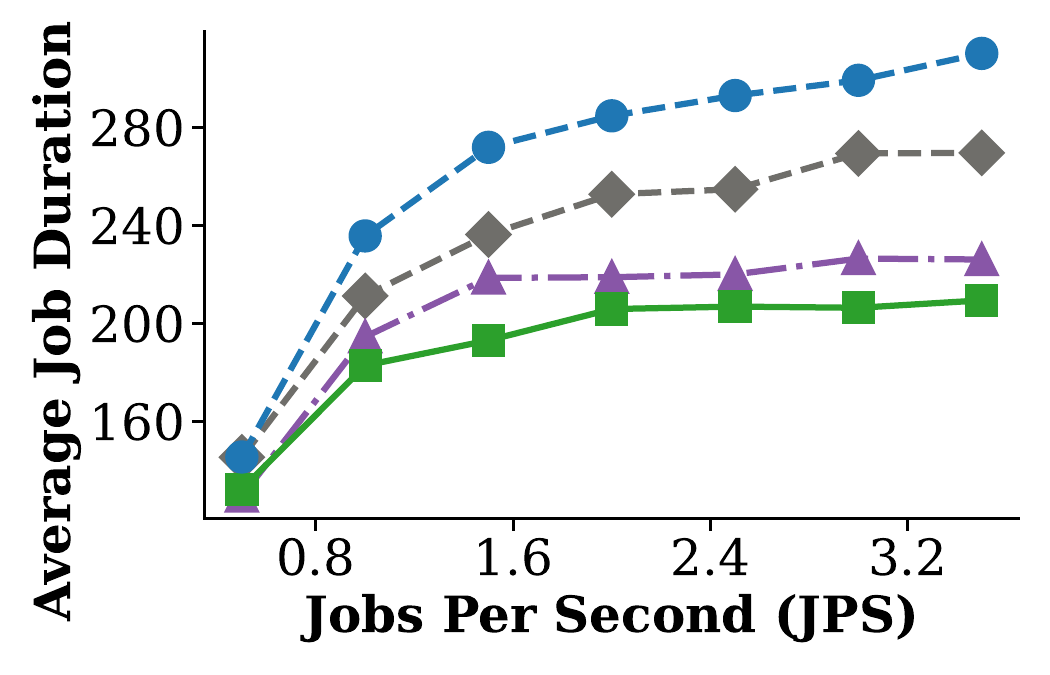}
        \subcaption{BFCL}
    \end{subfigure}
    \tightcaption{Contributions of individual ideas to \name. Program-level FCFS prioritizes requests with earlier program arrival instead of request arrival. Static TTL uses a fixed TTL threshold calculated from the cold-start handling mechanism.}
    \label{fig:queue_ablation}
\end{figure}
\vspace{1pt}

\mypara{Real SWE-Agent in Distributed Setting}
In order to fully evaluate \name's performance in real-world deployment scenarios at scale. 
We test \name running a real SWE agent for 500 tasks in SWE-Bench-Verified in \company's internal H100 testbed. 
We set up our agent client environment by adding a job distributor for the SWE-Bench platform that distributes agents according to a Poisson distribution.
We use a simple session-aware routing for \name and compare against other distributed inference solutions.
We measure the per-job finish time and collect the pass rate of each agent program for their generated results on SWE-bench after generation.

As demonstrated by Figure~\ref{fig:real_exp}, \name consistently outperforms baselines in terms of average delay when pass rates are equal. Notice that \name actually has higher pass rate than baselines. This is due to SWE-Bench's time limit for environment Docker containers to prevent hanging. When the baseline's running time exceeds 15 minutes it will be preempted and treated as a failure case. This proves \name's usability in real production settings.

\tightsubsection{Sensitivity Analysis}

\mypara{Varying Inference Engine Configuration}
In order to show that \name is robust to varying inference engine configurations, 
we evaluate \name with different configurations of the inference engine.
In Figure~\ref{fig:eval_batch_chunk}, we set the job per second to be 0.13 and vary the maximum batch size to compare \name with different baselines. 
As we can see, \name's improvement remains stable across different batch sizes.
Moreover, in Figure~\ref{fig:eval_batch_chunk}, we vary the number of chunk size from 256 to 4096. We observe similar improvements across different chunk sizes.
This demonstrates the robustness of our approach to different inference engine configurations.

\mypara{Scaling Law for Turn Numbers}
Figure \ref{fig:robust_study} evaluates our scheduler's robustness in multi-turn scenarios. 
We simulate more-turn scenarios on SWE-Bench by repeating the trace (1$\times$ to 5$\times$) while inversely scaling the token lengths to emulate more turns but make total token fit within the context window. 
With a request rate of 0.13 JPS and 200 GB for DRAM offloading, the results show that the baseline methods degrade as the number of turns increases. 
This is because the increased number of turns leads to more tool calls and longer overall execution times, exacerbating the scheduling challenges faced by traditional methods.
In contrast, our approach maintains stable, low-latency performance, demonstrating its effectiveness for complex, many-turn agentic interactions.

\mypara{SSD Offloading}
Similar to CPU offloading, SSD offloading offers bigger space but slower loading. We evaluate \name  with extended SSD storage layer beyond CPU offloading using LMCache on SWE-bench workload with llama-8B on B200.  As shown in Figure~\ref{fig:ssd}, \name consistently improves average delay compared with baselines when also utilizing disks of different sizes.

\tightsubsection{Ablation Studies and Microbenchmarking}

\mypara{Ablation Study}
We conduct an ablation study to analyze the impact of our cost modeling on \name's overall performance. 
In Figure~\ref{fig:queue_ablation}, we compare \name with baselines that only applies part of the optimizaions. Program-Level FCFS changes the original request-level FCFS in vLLM into priority based on program arrival. Static TTL builds upon program-level FCFS to utilize fixed TTL threshold estimated cold-start handling. As demonstrated, different ideas of \name gradually improves performance.

\mypara{Scheduler Overhead}
As shown in Table~\ref{tab:latency_schedule}, our approach introduces a minor scheduling overhead compared to the baselines. 
However, this overhead is on the order of single-digit milliseconds, 
which is negligible compared to the GPU execution time for LLM inference. 
The significant end-to-end performance improvements from our scheduling strategy far outweigh this small increase in scheduling latency.

\mypara{Application to Reinforcement Learning}
We also conducted a micro-benchmark for potential reinforcement learning use of \name. We tested the OpenHands Agent with GLM-4.5-fp8 training on Multi-SWE bench~\cite{zan2025multiswebenchmultilingualbenchmarkissue} for rollout generation. The hardware setup is an 8xH100 node.  We compared with the concurrent RL work ThunderAgent~\cite{kang2026thunderagentsimplefastprogramaware} on inference steps per minute, as reported by the original paper. As demonstrated by Table~\ref{tab:rl}, \name achieves higher throughput for single node rollout. 

\begin{table}[t]
\tighttablecaption{\name introduces minor scheduling latency overhead comparison under different DRAM offloading settings.}
\label{tab:latency_schedule}
\centering
\resizebox{0.8\linewidth}{!}{
    \begin{tabular}{lcc}
    \toprule
    \textbf{System} & \textbf{No CPU Offload} & \textbf{CPU Offload} \\
    \midrule
    vLLM      & 0.95\,ms & 2.33\,ms \\
    Autellix  & 0.82\,ms & 2.18\,ms \\
    InferCept & N/A      & 2.25\,ms \\
    Ours      & 0.96\,ms & 2.30\,ms \\
    \bottomrule
    \end{tabular}
}
% \vspace{10pt}
% \vspace{-12pt}

\end{table}

\begin{table}[t]
\tighttablecaption{\name achieves better performance on OpenHands rollout than concurrent work.}
\label{tab:rl}
\centering
\resizebox{\linewidth}{!}{
    \begin{tabular}{lccc}
    \toprule
     & \textbf{vLLM} & \textbf{ThunderAgent} & \textbf{Continnum} \\
    \midrule
    Throughput (Steps Per Min) & 93.4 & 114.8 & \textbf{144.9} \\
    \bottomrule
    \end{tabular}
}
\end{table}

\tightsection{Related Work}
% \vspace{-5pt}
\mypara{LLM Inference Systems} 
There have been many research papers on improving LLM inference. Serving engines including vLLM~\cite{kwon2023efficient} and SGLang~\cite{zheng2024sglang} achieves state of the art inference by adapting paged attention design and optimized kernels. Besides the wide range of kernel-level optimizations that improve GPU execution speed~\cite{ye2025flashinferefficientcustomizableattention,  dao2022flashattentionfastmemoryefficientexact, zhu2025nanoflowoptimallargelanguage}, researchers have also proposed many optimizations on resource management: continuous batching~\cite{yu2022orca}, chunked prefill~\cite{agrawal2024taming}, skip-join multi-level scheduling~\cite{wu2023fast}. Many of them have been ported into the inference engine. 
Previous work have also explored efficient offloading to CPU DRAM and disks~\cite{gao2024attentionstore, hicache, cheng2025lmcache, liu2024cachegen,yao2025cacheblend}.
For distributed inference, people have adopted session aware routing~\cite{srivatsa2024prebleefficientdistributedprompt, vllm-production-stack}, KV-cache aware routing~\cite{xia2025skywalkerlocalityawarecrossregionload}, and prefill-decode disaggregation~\cite{zhong2024distserve}. Building upon these work, \name extends LLM inference into long-horizon multi-turn agentic workloads and improves resource management when resources are competed by different requests.

\mypara{Time-to-live Mechanisms in Computer Systems} Time-to-live (TTL) is a longstanding abstraction in computer systems design, widely used in DNS resolvers, distributed caches, CDN edge nodes, and consistency protocols to bound staleness and prevent unbounded resource retention~\cite{krishnamurthy2001use, jung2003modeling, cohen2005performance, nishtala2013scaling, basu2018adaptive, moura2019cache, lawrence2020serving, yang2021large, hernandez2021reduction, hendri2024optimizing}. 
In these settings, TTL acts as a coarse-grained validity window that balances freshness, load, and robustness under unpredictable update or fetch latencies.
We build on this lineage but extend TTL to a new domain: fine-grained resource management inside LLM inference engines.
Unlike traditional TTL uses, where entries are independent and correctness constraints are semantic rather than performance-critical, KV caches interact tightly with GPU memory pressure, prefill costs, and scheduling fairness in LLM serving engines.
% \name adapts TTL to this environment by modeling the benefit of preserving multi-turn continuity for agent workflows and the cost of blocking future requests in shared inference engines.
To our knowledge, \name is the first system to use TTL to regulate LLM KV cache as a function of predicted tool-call durations, scheduling-side delay propagation, and workload pattern.

% It specifies a finite period for which a piece of information is considered valid or cached before it must be refreshed or discarded. 
% For instance, in the Domain Name System (DNS), TTL values dictate how long a resolver should cache a DNS record, balancing the load on authoritative servers with the need for timely propagation of updates~\cite{moura2019cache, lawrence2020serving, hernandez2021reduction}. 
% Similarly, Content Delivery Networks (CDNs) and web caches use TTL to determine how long to store and serve content from edge locations before re-fetching it from the origin server~\cite{basu2018adaptive, krishnamurthy2001use, hendri2024optimizing}.

% The primary goal of TTL is to strike a balance between performance and data consistency. 
% A short TTL ensures that clients receive fresh data but can increase traffic to the origin, 
% while a long TTL reduces origin load and improves response times at the cost of potential data staleness. 
% Setting an appropriate TTL is therefore a critical configuration choice that depends on the specific application's tolerance for stale data and its performance requirements. 
% Our work adapts this fundamental concept to the domain of LLM agent inference, using TTL to manage the lifecycle of KV caches for agentic workloads.
\vspace{-2pt}
\mypara{Generality Beyond ReAct-Style Agents}
The current design of \name are optimized for ReAct-style, tool-interleaving agents where each LLM step returns a clear tool invocation followed by a gap before the next step.
\name naturally extends to parallel tool calls since it still follows the 
sequential “reason -> tool -> reason” rhythm.
Some emerging agent frameworks, however, could involve non-linear control flows: speculative branches, asynchronous multi-agent coordination, and context folding. Although such workloads are mostly experimental and yet to be tested in real production workloads, their inference pattern may violate the sequential flow and requires future change. 
Extending \name to support such workloads is an important direction for future work.

\tightsection{Conclusion}

Agentic workloads introduce new scheduling challenges for LLM serving systems due to frequent tool calls, highly variable inter-step delays, and the need to preserve multi-turn continuity. 
We present \name, a KV cache retention and scheduling system that balances both the benefit of cache reuse and the cost of blocking GPU memory through a time-to-live mechanism. 
By integrating TTL-based pinning with program-level FCFS , \name reduces unnecessary prefills, mitigates per-turn queueing delays, and robustly adapts to unpredictable tool-call latencies. 
Our implementation on top of vLLM shows consistent improvements in end-to-end job completion time across model sizes, hardware configurations, and real-world agent workloads. 
\name demonstrates that principled, tool-aware KV management is essential for efficient multi-turn agent serving. We hope it lays the groundwork for future systems to deeply integrate agent workload into LLM inference engines.

\section{Data Availability}
The implementation and artifacts are available at: \noindent\url{https://github.com/Hanchenli/vllm-continuum}

% We present Continnum, a KV-cache retention and scheduling system for multi-turn agent workloads. Continnum uses TTL-based pinning to retain KV cache across tool calls while accounting for cache miss cost and per-turn queueing delay. Combined with program-level FCFS scheduling, it reduces unnecessary prefills, preserves execution continuity, and adapts to variable tool latencies. Across models, hardware platforms, and real agent workloads, Continnum consistently improves end-to-end job completion time, highlighting the importance of tool-aware KV-cache management for agent serving.

\clearpage
\pagestyle{plain}

% \markboth{}{} 
% \renewcommand{\shortauthors}{}
% \renewcommand{\shorttitle}{}

% \nocite{*}
\bibliographystyle{ACM-Reference-Format}
\bibliography{references}
% Appendix material is kept in sections/appendix.tex but omitted from the main
% VLDB submission PDF to keep the paper within the 12-page body limit.
% \clearpage
% \appendix
% \input{sections/appendix}
%%%%%%%%%%%%%%%%%%%%%%%%%%%%%%%%%%%%%%%%%%%%%%%%%%%%%%%%%%%%%%%%%%%%%%%%%%%%%%%%
\end{document}